\documentclass[nofootinbib,aps,a4paper,letterpaper,superscriptaddress,
onecolumn,showpacs,times,eqsecnum]{revtex4}
\usepackage{amsmath}
\usepackage{amsfonts}
\usepackage{booktabs}
\usepackage{multirow}
\usepackage{siunitx} 
\usepackage{adjustbox}
\pdfoutput=1
\usepackage{graphicx}
\usepackage{subcaption}
\usepackage{ulem}
\usepackage{braket}
\usepackage{dcolumn}
\usepackage{bm,url}

\usepackage[dvipsnames,svgnames]{xcolor}  
\usepackage[linktocpage]{hyperref}
\definecolor{oxfordblue}{rgb}{0.0, 0.13, 0.28}
\definecolor{burgundy}{rgb}{0.5, 0.0, 0.13}
\definecolor{darkolivegreen}{rgb}{0.33, 0.42, 0.18}
\definecolor{darkblue}{rgb}{0,0,0.5}
\definecolor{richcarmine}{rgb}{0.84, 0.0, 0.25}
\definecolor{darkblue}{rgb}{0,0,0.5}
\definecolor{bluer}{rgb}{0.00,0.50,0.75}{}
\hypersetup{colorlinks=true, citecolor=red, linkcolor=blue,
 urlcolor = magenta, filecolor=magenta}

\usepackage{ragged2e}

\makeatletter
\long\def\@makecaption#1#2{%
  \par\small
  \begingroup
  \justifying
  \noindent #1.\ #2\par
  \endgroup
}
\makeatother

\begin{document}
 
 \newcommand\be{\begin{equation}}
  \newcommand\ee{\end{equation}}
 \newcommand\bea{\begin{eqnarray}}
  \newcommand\eea{\end{eqnarray}}
 \newcommand\bseq{\begin{subequations}} 
  \newcommand\eseq{\end{subequations}}
 \newcommand\bcas{\begin{cases}}
  \newcommand\ecas{\end{cases}}
 \newcommand{\p}{\partial}
 \newcommand{\f}{\frac}
 
\title{Higgs-like inflation in scalar-torsion $f(T,\phi)$ gravity in light of ACT-SPT-DESI constraints}

\author{Nitesh Kumar\,\href{https://orcid.org/0009-0005-4716-0663}{%
\includegraphics[width=8pt]{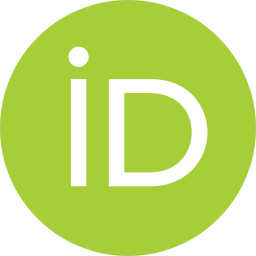}}}
\email{nitesh.kumar@postgrado.uv.cl}
\affiliation{Instituto de Física y Astronomía, Facultad de Ciencias, Universidad de Valparaíso, Gran Bretaña 1111, Valparaíso, Chile.}
\affiliation{Departamento de F\'isica, Facultad de Ciencias, Universidad de
Tarapac\'a, Casilla 7-D, Arica, Chile.}
\affiliation{Departamento de Física, Facultad de Ciencias, Universidad de La Serena, Avenida Cisternas 1200, La Serena, Chile.}
\author{Giovanni Otalora\,\href{https://orcid.org/0000-0001-6753-0565}{%
\includegraphics[width=8pt]{orcid.png}}}
\email{giovanni.otalora@academicos.uta.cl}
\affiliation{Departamento de F\'isica, Facultad de Ciencias, Universidad de
Tarapac\'a, Casilla 7-D, Arica, Chile.}
\author{Rodrigo Reyes\,\href{https://orcid.org/0009-0000-3569-5728}{%
\includegraphics[width=8pt]{orcid.png}}}
\email{ rodrigo.reyes.pizarro@alumnos.uta.cl}
\affiliation{Departamento de F\'isica, Facultad de Ciencias, Universidad de
Tarapac\'a, Casilla 7-D, Arica, Chile.}
\author{Bastian Espinoza\,\href{https://orcid.org/0009-0007-2093-6112}{%
\includegraphics[width=8pt]{orcid.png}}}
\email{bastian.espinoza.leal@alumnos.uta.cl }
\affiliation{Departamento de F\'isica, Facultad de Ciencias, Universidad de
Tarapac\'a, Casilla 7-D, Arica, Chile.}
\author{Manuel Gonzalez-Espinoza\,\href{https://orcid.org/0000-0003-0961-8029}{%
\includegraphics[width=8pt]{orcid.png}}}
\email{manuel.gonzalez@pucv.cl}
\affiliation{Instituto de F\'{\i}sica, Pontificia Universidad Cat\'olica de Valpara\'{\i}so, Casilla 4950, Valpara\'{\i}so, Chile}
\author{Emmanuel N. Saridakis\,\href{https://orcid.org/0000-0003-1500-0874}{%
\includegraphics[width=8pt]{orcid.png}}}
\email{msaridak@noa.gr}
 \affiliation{Institute for Astronomy, Astrophysics, Space Applications and 
Remote Sensing, National Observatory of Athens, 15236 Penteli, Greece}
\affiliation{CAS Key Laboratory for Research in Galaxies and Cosmology,  
 University of Science and Technology of China, Hefei, Anhui 230026,
China}
\affiliation{Departamento de Matem\'{a}ticas, Universidad Cat\'{o}lica del Norte, 
Avda.
Angamos 0610, Casilla 1280 Antofagasta, Chile}

\begin{abstract}
We study Higgs-like inflation in the framework of scalar-torsion gravity, focusing on the general class of $f(T,\phi)$ theories in which gravitation is mediated by torsion rather than curvature. Motivated by the increasing precision of cosmic microwave background and large-scale-structure observations, we examine whether Higgs-like inflation remains compatible with current data in this extended gravitational setting. Working within the slow-roll approximation, we analyze the inflationary dynamics both analytically and numerically. In the dominant-coupling regime we derive closed-form expressions for the scalar spectral index and the tensor-to-scalar ratio as functions of the number of e-folds, and we subsequently relax this assumption by numerically solving the slow-roll equations. Confrontation with the latest constraints from Planck 2018, ACT DR6, DESI DR1, and BICEP/Keck shows that Higgs-like inflation in $f(T,\phi)$ gravity is fully consistent with current bounds, naturally accommodating the preferred shift in the scalar spectral index and leading to distinctive tensor-sector signatures.
\end{abstract}

\maketitle

\section{Introduction}

The inflationary paradigm, which is based on a brief period of accelerated
expansion preceding the standard Hot Big Bang evolution, constitutes one of
the cornerstones of modern cosmology. This early phase of quasi-exponential
expansion provides a compelling resolution to several conceptual problems of
the standard cosmological model, including the flatness, horizon, and monopole
problems~\cite{Starobinsky,Guth:1980zm,Albrecht:1982wi,Linde:1981mu}. Moreover,
inflation offers a natural mechanism for generating the primordial
inhomogeneities that seeded the cosmic microwave background (CMB) anisotropies
and the formation of large-scale structure (LSS). These inhomogeneities are
believed to originate from quantum fluctuations amplified to cosmological
scales during inflation, resulting in a nearly scale-invariant, adiabatic, and
Gaussian spectrum of curvature 
perturbations~\cite{Weinberg:2008zzc,Baumann:2018muz}.

A key observational probe of inflationary dynamics is the scalar spectral index
$n_s$, which characterizes the scale dependence of the primordial curvature
power spectrum. The Planck 2018 analysis reported $n_s=0.9649\pm0.0042$ (68\%
C.L.)~\cite{Planck:2018jri}, together with an upper bound on the 
tensor-to-scalar
ratio, $r_{0.05}<0.036$, from BICEP/Keck~\cite{BICEPKeck:2021gln}. More recent
high-precision measurements have further sharpened these constraints. The
combination of Planck with South Pole Telescope (SPT) data yields
$n_s=0.9647\pm0.0037$~\cite{SPT-3G:2024atg}, while the latest Atacama Cosmology
Telescope (ACT) results combined with Planck shift the preferred value upward
to 
$n_s=0.9709\pm0.0038$~\cite{AtacamaCosmologyTelescope:2025blo,
AtacamaCosmologyTelescope:2025nti}.
Including DESI DR1 large-scale structure data further tightens the constraint
to $n_s=0.9743\pm0.0034$~\cite{AtacamaCosmologyTelescope:2025blo}, notably 
closer
to exact scale invariance. This steady improvement in precision reveals the
increasing distinguishing power of current and upcoming cosmological surveys.

Beyond the standard observables $n_s$ and $r$, inflation generically predicts
additional signatures. A stochastic background of primordial gravitational
waves (PGWs), if detected, would provide a direct window into the energy scale
of inflation and the quantum nature of gravity in the early Universe
\cite{Kamionkowski:2015yta,Meerburg:2015zua}. Current bounds are dominated by
BICEP/Keck~\cite{BICEPKeck:2021gln}, while forthcoming CMB polarization
experiments, such as the Simons Observatory~\cite{SimonsObservatory:2018koc},
CMB-S4~\cite{CMB-S4:2016ple}, and LiteBIRD~\cite{LiteBIRD:2022cnt}, aim to reach
sensitivities of $\sigma(r)\sim10^{-3}$. Complementary probes will be provided
by space-based interferometers, including
LISA~\cite{LISACosmologyWorkingGroup:2022jok} and proposed missions such as
DECIGO~\cite{Kawamura:2020pcg}, which explore a broader frequency range and are
sensitive to gravitational waves generated during or after inflation.

Inflation may also enhance curvature perturbations on small scales, potentially
leading to the formation of primordial black holes (PBHs). PBHs remain viable
dark-matter candidates and are constrained by a wide array of observations,
including gravitational-wave detections by LIGO/Virgo/KAGRA
\cite{LIGOScientific:2021job}, microlensing surveys such as Subaru/HSC, OGLE, 
and
EROS~\cite{Niikura:2019kqi,Takhistov:2020vxs}, and their effects on the CMB and
the high-redshift 21-cm signal~\cite{Clark:2016nst,Ali-Haimoud:2016mbv}. Future
facilities, including the Roman Space Telescope~\cite{Fardeen:2023euf} and the
Square Kilometre Array (SKA)~\cite{Mena:2019nhm,Zhao__2026}, are expected to
further tighten these constraints and probe a wide PBH mass range.

The steadily improving constraints on $n_s$ and $r$, particularly those
incorporating ACT and DESI data, have renewed interest in reassessing the
viability of leading inflationary scenarios. Plateau-type models, such as the
Starobinsky $R^{2}$ model~\cite{Starobinsky} and $\alpha$-attractor
constructions~\cite{Kallosh:2013hoa,Kallosh:2013yoa,Kallosh_2021}, have long
been favored by Planck-only analyses. However, the upward shift in the preferred
value of $n_s$ places these scenarios under increasing tension, with the
Starobinsky model approaching exclusion at the $2\sigma$ level
\cite{Linde:2025pvj,Kallosh:2025ijd}, a situation that holds for many scalar-tensor models too
 \cite{Tsujikawa:2014mba, Germani:2015plv,  BeltranJimenez:2017cbn, Sebastiani:2017cey,
 Oikonomou:2020sij,  Chen:2021nio,Oikonomou:2021hpc}. Conversely, models previously disfavored
re-emerge as viable candidates, including certain monomial potentials and
quadratic inflation supplemented by non-minimal gravitational couplings
\cite{Roest:2013fha,Kallosh:2025ijd,Kallosh:2025rni}. This evolving observational landscape
motivates the exploration of alternative inflationary frameworks, including
modified gravity scenarios that can accommodate the data while offering new
phenomenology.

In this context, the Higgs field stands out as a particularly compelling
inflaton candidate, being the only experimentally confirmed fundamental scalar
field within the Standard Model \cite{ATLAS:2012yve,CMS:2012qbp}. In its minimal
form, however, the Higgs quartic potential is too steep to support slow-roll
inflation. Introducing a non-minimal coupling to gravity effectively flattens
the potential at high energies, leading to successful Higgs inflation and
establishing a direct link between particle physics and early-Universe
cosmology~\cite{Fakir:1990eg,Bezrukov:2007ep}. This idea has motivated extensive
work, including effective-field-theory 
treatments~\cite{Bezrukov:2010jz,Tronconi:2025gmn},
LHC-driven constraints~\cite{Wang:2021ayg}, non-minimal derivative coupling modifications \cite{Germani:2010gm,Karydas:2021wmx},
and recent extensions inspired by
ACT, SPT, and DESI data~\cite{Han:2025cwk,6fpc-67s1,Aoki:2025wld,Roy:2026vwx}. These
developments naturally motivate the study of Higgs inflation within
alternative gravitational frameworks.

Among such alternatives, modified gravity theories based on torsion rather
than curvature have attracted growing interest~\cite{Cai:2015emx}. These
theories generalize the Teleparallel Equivalent of General Relativity (TEGR),
originally proposed by Einstein~\cite{Einstein,TranslationEinstein}, by
describing gravity through spacetime torsion instead of curvature. Prominent
extensions include $f(T)$ gravity
\cite{Ferraro:2006jd,Linder:2010py,Harko:2014sja,Harko:2014aja,Krssak:2015oua,Leyva:2021fuo}, scalar-torsion
theories
\cite{Geng:2011aj,Geng:2011ka,Xu:2012jf,Otalora:2013tba,Otalora:2013dsa,
Otalora:2014aoa,Skugoreva:2014ena,Hohmann:2018rwf,Gonzalez-Espinoza:2020azh,Gonzalez-Espinoza:2020jss,Gonzalez-Espinoza:2021mwr,Gonzalez-Espinoza:2021qnv,Rodriguez-Benites:2024pce,Duchaniya:2022hiy,Duchaniya:2022fmc,Villalobos-Silva:2026deu},
and higher-order torsional modifications
\cite{Otalora:2016dxe,Gonzalez-Espinoza:2021nqd}. In scalar-torsion models, a
scalar field couples nontrivially to torsion, leading to distinctive
inflationary dynamics and potentially observable imprints in the primordial
perturbation spectra. The role of local Lorentz symmetry breaking and its
impact on cosmological perturbations has been explored in this context
\cite{Gonzalez-Espinoza:2020azh}, and reconstruction techniques for the
inflationary observables $n_s(N)$ and $r(N)$ have also been developed
\cite{Gonzalez-Espinoza:2021qnv}.

In this work, we revisit Higgs-like inflation within the framework of scalar-torsion gravity described by a general function $f(T,\phi)$, with $\phi$ identified as a Higgs-like field. While Higgs-like inflation in torsion-based theories has previously been considered only in more restricted settings~\cite{Raatikainen:2019qey}, here we extend the analysis to the full scalar-torsion $f(T,\phi)$ framework. To place our results in context, we note that inflation in scalar-torsion gravity has been investigated in only a limited number of studies, and Higgs-like realizations are even scarcer. Although Ref.~\cite{Raatikainen:2019qey} explored Higgs inflation in this setting, it did not present a perturbation analysis within the scalar-torsion framework. More general scalar-torsion inflationary scenarios with nonminimal couplings and Galileon-type interactions were considered in Ref.~\cite{Gonzalez-Espinoza:2019ajd}, but without implementing the scalar and tensor power spectra that arise once the additional degrees of freedom inherent to the theory are properly taken into account. Several related works~\cite{Chakrabortty2021Inflation,Chakrabortty2022Correction,Fomin2025ScalarTorsion} instead evaluate observables using GR-based scalar-tensor power spectra and consistency relations, which are not valid in scalar-torsion gravity because the perturbation sector is modified relative to GR. By contrast, our analysis derives inflationary observables from the scalar and tensor power spectra in scalar-torsion gravity developed in Refs.~\cite{Gonzalez-Espinoza:2020azh,Gonzalez-Espinoza:2021qnv}, where the perturbation sector is treated consistently and the effects of local Lorentz-symmetry breaking are properly included. We then confront the resulting predictions with current data. For completeness, Higgs-like inflation has also been explored in the Einstein-Cartan framework~\cite{Shaposhnikov2021Erratum,Piani_2022,Rigouzzo_2022}.  However, torsion in Einstein-Cartan theory is associated with a different geometric structure than in teleparallel gravity, so the corresponding inflationary dynamics and observable predictions are not directly comparable.

Building on these results, and making use of the general expressions for the scalar and tensor primordial power spectra derived in Ref.~\cite{Gonzalez-Espinoza:2020azh}, we perform both an analytical and a numerical study of the inflationary dynamics. In the dominant-coupling regime, we derive closed-form expressions for the inflationary observables as functions of the number of e-folds $N$. Beyond this regime, we carry out a numerical analysis within the slow-roll approximation, without assuming strong coupling. Finally, we confront the predictions of the model with the latest CMB and LSS constraints, including Planck~2018 \cite{Planck:2018jri}, ACT~DR6 \cite{AtacamaCosmologyTelescope:2025blo}, DESI~DR1 \cite{Adame_2025}, and BICEP/Keck~2018 \cite{BICEPKeck:2021gln}, in order to assess the viability of Higgs-like inflation in $f(T,\phi)$ gravity and to constrain its parameter space.

The paper is organized as follows. In Section \ref{Themodel} we present the 
theoretical
framework of scalar-torsion $f(T,\phi)$ gravity, including the cosmological
background and the formulation of linear perturbations. In Section 
\ref{Inflationarydynamics} we
analyze the inflationary dynamics, introducing the slow-roll formulation and
deriving analytical results in the dominant-coupling regime. Section 
\ref{AnalyticalResults} is
devoted to Higgs-like inflation, where we obtain explicit analytical predictions
for the inflationary observables and confront them with observational data.
In Section \ref{beyondhighenergy} we perform a numerical analysis of Higgs-like 
inflation beyond the
dominant-coupling regime. Finally, we summarize our results and present our
conclusions in Section \ref{Conclusions}.

\section{Scalar-torsion \texorpdfstring{$f(T,\phi)$}{f(T,φ)} gravity and cosmology}
\label{Themodel}

In this section we present the theoretical framework underlying our analysis.
We briefly review teleparallel gravity as a torsion-based formulation of
gravitational interactions, and then we introduce its scalar-torsion extension
described by a general function $f(T,\phi)$, where $T$ denotes the torsion 
scalar
and $\phi$ is a dynamical scalar field. We subsequently focus on a
homogeneous and isotropic cosmological background and we derive the equations
governing the background evolution and linear perturbations relevant for
inflationary dynamics. This formulation provides the basis for the study of
slow-roll inflation, the derivation of primordial scalar and tensor spectra,
and the confrontation of the model with current cosmological observations
carried out in the following sections.

\subsection{Teleparallel gravity}
\label{Intro_TG}

Teleparallel gravity (TG) provides an alternative geometric formulation of
gravitational interactions, which can be understood as a gauge theory for the
translation group $T_{4}$ 
\cite{Early-papers5,Early-papers6,Aldrovandi-Pereira-book,Pereira:2019woq}.
In this approach spacetime is described by a principal fiber bundle
$P(M,T_{4})$, with base manifold $M$ and structure group $T_{4}$
\cite{Aldrovandi-Pereira-book,JGPereira2,Arcos:2005ec}.
The fundamental gravitational variable is the tetrad field
$e^{A}{}_{\mu}$, which arises as the gauge potential associated with local
translations. It provides the soldering between the tangent space and the base
manifold and defines the spacetime coframe.
The spacetime metric is constructed from the tetrads as
\begin{equation}
g_{\mu\nu} = \eta_{AB}\, e^{A}{}_{\mu} e^{B}{}_{\nu},
\end{equation}
where $\eta_{AB}=\mathrm{diag}(-1,1,1,1)$ is the Minkowski metric.

In TG the Lorentz (spin) connection $\omega^{A}{}_{B\mu}$ is interpreted as the
local representative of a global principal connection on the Lorentz bundle
$P(M,SO(1,3))$.
If $\omega$ denotes the Lorentz-valued connection one-form on the principal
bundle and $\sigma:U\subset M\to P(M,SO(1,3))$ is a local section, the spacetime components
of the spin connection are obtained through the pullback
\begin{equation}
\omega^{A}{}_{B\mu} = (\sigma^{*}\omega)^{A}{}_{B\mu}.
\end{equation}
A defining feature of teleparallelism is that this Lorentz connection is chosen
to be flat, encoding only inertial effects.
Accordingly, it can always be written in the form
\begin{equation}
\omega^{A}{}_{B\mu}
 = \Lambda^{A}{}_{D}(x)\,\partial_{\mu}\Lambda_{B}{}^{D}(x),
\end{equation}
with $\Lambda^{A}{}_{B}(x)\in SO(1,3)$, and its curvature tensor identically
vanishes, namely
\begin{equation}
R^{A}{}_{B\mu\nu}
 = \partial_{\mu}\omega^{A}{}_{B\nu}
   - \partial_{\nu}\omega^{A}{}_{B\mu}
   + \omega^{A}{}_{C\mu}\omega^{C}{}_{B\nu}
   - \omega^{A}{}_{C\nu}\omega^{C}{}_{B\mu}
 = 0.
\end{equation}

Gravitational effects are instead encoded in the torsion tensor, constructed
from the tetrad and the spin connection.
In particular, torsion plays the role of the field strength associated with local translations,
in close analogy with the
field strength of Yang-Mills theories, and it is given by
\begin{equation}
T^{A}{}_{\mu\nu}
= \partial_{\mu} e^{A}{}_{\nu}
 - \partial_{\nu} e^{A}{}_{\mu}
 + \omega^{A}{}_{B\mu} e^{B}{}_{\nu}
 - \omega^{A}{}_{B\nu} e^{B}{}_{\mu}.
\end{equation}

Using the tetrad, one introduces the Weitzenb\"ock connection as
\begin{equation}
\Gamma^{\rho}{}_{\nu\mu}
= e_{A}{}^{\rho}\,\partial_{\mu}e^{A}{}_{\nu}
+ e_{A}{}^{\rho}\,\omega^{A}{}_{B\mu} e^{B}{}_{\nu},
\end{equation}
which is metric-compatible and curvature-free, but possesses nonvanishing
torsion.
It differs from the Levi-Civita connection $\bar{\Gamma}^{\rho}{}_{\nu\mu}$ by
the contortion tensor
\begin{equation}
K^{\rho}{}_{\nu\mu}
= \frac12 \left(
T_{\nu}{}^{\rho}{}_{\mu}
+ T_{\mu}{}^{\rho}{}_{\nu}
- T^{\rho}{}_{\nu\mu}
\right),
\end{equation}
such that
\begin{equation}
\Gamma^{\rho}{}_{\nu\mu}
= \bar{\Gamma}^{\rho}{}_{\nu\mu} + K^{\rho}{}_{\nu\mu}.
\end{equation}

The teleparallel action has a gauge-theoretic structure and is constructed from
quadratic invariants of the torsion tensor. Now, the fundamental scalar quantity is the torsion scalar
\begin{equation}
T = S_{\rho}{}^{\mu\nu} T^{\rho}{}_{\mu\nu},
\label{torsionsclara}
\end{equation}
where the superpotential is defined as
\begin{equation}
S_{\rho}{}^{\mu\nu}
= \frac12\left(
K^{\mu\nu}{}_{\rho}
+ \delta^{\mu}{}_{\rho} T^{\theta\nu}{}_{\theta}
- \delta^{\nu}{}_{\rho} T^{\theta\mu}{}_{\theta}
\right).
\end{equation}
Hence, the action of teleparallel gravity   reads
\begin{equation}
S = -\frac{1}{2\kappa^{2}} \int d^{4}x\, e\, T,
\end{equation}
where $e=\det(e^{A}{}_{\mu})=\sqrt{-g}$.

It can be shown that the torsion scalar and the Ricci scalar $\bar{R}$ 
constructed
from the Levi-Civita connection differ only by a total divergence, namely 
\begin{equation}
T = -\bar{R} + 2 e^{-1}\partial_{\mu}\bigl(e\,T_{\nu}{}^{\mu\nu}\bigr),
\end{equation}
implying that teleparallel gravity is dynamically equivalent to general
relativity at the level of the field equations. This formulation is therefore
referred to as the Teleparallel Equivalent of General Relativity (TEGR).

This geometric framework provides a natural starting point for constructing
torsion-based extensions of gravity.
Cosmological models involving scalar fields nonminimally coupled to torsion have
been extensively investigated
\cite{Geng:2011aj,Otalora:2013tba,Otalora:2013dsa,Otalora:2014aoa,
Skugoreva:2014ena},
while nonlinear generalizations such as $f(T)$ gravity
\cite{Bengochea:2008gz,Linder:2010py} introduce genuinely new gravitational
dynamics with no curvature-based counterpart
\cite{Li:2011wu,Gonzalez-Espinoza:2018gyl}.
These developments motivate the study of more general scalar-torsion theories
of the form $f(T,\phi)$ \cite{Hohmann:2018rwf,Gonzalez-Espinoza:2020azh,Gonzalez-Espinoza:2020jss,Gonzalez-Espinoza:2021mwr,Gonzalez-Espinoza:2021qnv,Rodriguez-Benites:2024pce,Duchaniya:2022hiy,Duchaniya:2022fmc, Villalobos-Silva:2026deu}, as well as teleparallel analogues of Horndeski-like scalar-tensor models \cite{Bahamonde:2019shr}, which will be the focus of the following subsection.

\subsection{Scalar-torsion \texorpdfstring{$f(T,\phi)$}{f(T,φ)} gravity}
\label{f(T,phi)}

Building on the teleparallel formulation outlined above, one can construct
modified gravity theories by allowing for nonlinear functions of the torsion
scalar and by introducing nontrivial couplings between torsion and additional
degrees of freedom.
In this work we focus on scalar-torsion theories described by a general
function $f(T,\phi)$, where the scalar field $\phi$ will later be identified
with the inflaton. Such models provide a broad and flexible framework for
studying early-Universe dynamics beyond general relativity.

We consider the action \cite{Gonzalez-Espinoza:2020azh}
\begin{equation}
S=\int d^{4}x\,e\,\left[f(T,\phi)+P(\phi)\,X\right],
\label{action1}
\end{equation}
where $f(T,\phi)$ is an arbitrary function of the torsion scalar $T$ and the
scalar field $\phi$, and
$X=-\partial_{\mu}\phi\,\partial^{\mu}\phi/2$ denotes the canonical kinetic
term. The function $P(\phi)$ allows for a non-canonical normalization of the
scalar-field kinetic sector.

Variation of the action with respect to the tetrad $e^{A}{}_{\mu}$ yields the
field equations
\begin{eqnarray}
&& f_{,T}\,G_{\mu\nu}
+ S_{\mu\nu}{}^{\rho}\,\partial_{\rho} f_{,T}
+ \frac{1}{4} g_{\mu\nu}\left(f - T f_{,T}\right)+
 \frac{P(\phi)}{4}\left(g_{\mu\nu} X
+ \partial_{\mu}\phi\,\partial_{\nu}\phi\right)=0,
\label{fieldeq-general}
\end{eqnarray}
where a comma denotes partial differentiation,
$f_{,T}\equiv \partial f/\partial T$, and
$S_{\mu\nu}{}^{\rho}$ is the superpotential tensor defined in the previous
subsection.

The tensor $G_{\mu\nu}$ appearing in Eq.~\eqref{fieldeq-general} is the
teleparallel analogue of the Einstein tensor, and can be written in a
coordinate basis as
\begin{equation}
G^{\mu}{}_{\nu}
= e_{A}{}^{\mu}\,G^{A}{}_{\nu},
\end{equation}
with
\begin{equation}
G_{A}{}^{\mu}
\equiv e^{-1}\partial_{\nu}\!\left(e\,e_{A}{}^{\sigma}
S_{\sigma}{}^{\mu\nu}\right)
- e_{A}{}^{\sigma} T^{\lambda}{}_{\rho\sigma}
S_{\lambda}{}^{\rho\mu}
+ e_{B}{}^{\lambda} S_{\lambda}{}^{\rho\mu}
\omega^{B}{}_{A\rho}
+ \frac{1}{4} e_{A}{}^{\mu} T,
\end{equation}
as given in \cite{Aldrovandi-Pereira-book}.

The above equations generalize the Teleparallel Equivalent of General
Relativity (TEGR) to an arbitrary scalar-torsion interaction $f(T,\phi)$,
supplemented by a potentially non-canonical scalar kinetic term governed by
$P(\phi)$. This framework will serve as the starting point for our cosmological
analysis and the study of inflationary dynamics in the following subsections.

\subsection{Cosmological background and linear perturbations}

Let us now apply  $f(T,\phi)$ gravity at a cosmological framework, both at the background and perturbative levels. 

\subsubsection{Background evolution}

We begin by considering the homogeneous and isotropic cosmological background
relevant for inflationary dynamics. In particular, we impose the standard flat
Friedmann-Robertson-Walker (FRW) geometry by choosing the diagonal (proper) tetrad
\begin{equation}
e^{A}{}_{\mu}=\mathrm{diag}(1,a,a,a),
\label{veirbFRW}
\end{equation}
and the zero spin connection $\omega^{A}_{~~B \mu}=0$ \cite{Krssak:2015oua}, which corresponds to the spacetime metric
\begin{equation}
ds^{2}=-dt^{2}+a^{2}\,\delta_{ij}\,dx^{i}dx^{j},
\label{FRWMetric}
\end{equation}
where $a(t)$ is the scale factor and $t$ denotes cosmic time.

Substituting this background ansatz into the general field equations \eqref{fieldeq-general} we obtain the two Friedmann equations
\begin{eqnarray}
\label{Fr1}
f(T,\phi) - P(\phi) X - 2T f_{,T} &=& 0,
\label{phi} \\
f(T,\phi) + P(\phi) X - 2T f_{,T}
- 4\dot{H} f_{,T} - 4H \dot{f}_{,T} &=& 0, 
\label{Fr2}
\end{eqnarray}
as well as the Klein-Gordon equation
\begin{eqnarray}
- P_{,\phi} X - 3P(\phi)H\dot{\phi}
- P(\phi)\ddot{\phi} + f_{,\phi} &=& 0,
\label{KGequation}
\end{eqnarray}
where $H\equiv \dot{a}/a$ is the Hubble parameter, a dot denotes differentiation
with respect to $t$, and a comma indicates partial differentiation with respect
to $T$ or $\phi$. Additionally, note that inserting the FRW tetrad (\ref{veirbFRW}) into the torsion scalar (\ref{torsionsclara})
we find the useful expression 
$T=6H^2$.

Following Ref.~\cite{Gonzalez-Espinoza:2020azh}, it is convenient to express the
inflationary dynamics in terms of generalized slow-roll parameters. In
particular, one finds that the first slow-roll parameter can be written as
\begin{equation}
\epsilon = \delta_{PX} + \delta_{f_{,T}},
\label{slow_roll_eps}
\end{equation}
where we have introduced
\begin{eqnarray}
\epsilon &=& -\frac{\dot{H}}{H^{2}}, \nonumber \\
\delta_{PX} &=& -\frac{P(\phi) X}{2H^{2} f_{,T}}, \\
\delta_{f_{,T}} &=& \frac{\dot{f}_{,T}}{H f_{,T}} .
\label{slowpara1}
\end{eqnarray}
The parameter $\delta_{f_{,T}}$ quantifies deviations from the TEGR limit and
can be further decomposed as
\begin{equation}
\delta_{f_{,T}}=\delta_{f\dot{H}}+\delta_{fX},
\label{deltafT}
\end{equation}
with
\begin{equation}
\delta_{f\dot{H}}=\frac{f_{,TT}\dot{T}}{H f_{,T}},
\qquad
\delta_{fX}=\frac{f_{,T\phi}\dot{\phi}}{H f_{,T}} .
\end{equation}

Combining Eqs.~\eqref{slow_roll_eps} and \eqref{slowpara1}, one can express 
these
quantities in a compact form as
\begin{equation}
\delta_{f\dot{H}}
=-\frac{2\mu}{1+2\mu}\left(\delta_{PX}+\delta_{fX}\right),
\end{equation}
\begin{equation}
\delta_{f_{,T}}
=\frac{1}{1+2\mu}\left(\delta_{fX}-2\mu\,\delta_{PX}\right),
\end{equation}
and consequently
\begin{equation}
\epsilon
=\frac{1}{1+2\mu}\left(\delta_{PX}+\delta_{fX}\right),
\end{equation}
where we have defined the dimensionless deviation parameter
\begin{equation}
\mu\equiv \frac{T f_{,TT}}{f_{,T}},
\end{equation}
in analogy with the deviation parameter commonly used in curvature-based
modified gravity theories \cite{DeFelice:2010aj}.

During slow-roll inflation, the time variation of the slow-roll parameters is
suppressed, such that
$\dot{\delta}_{PX}\sim \dot{\delta}_{f_{,T}}
\sim \dot{\delta}_{fX}\sim\mathcal{O}(\epsilon^{2})$,
and similarly for higher-order parameters. These relations justify the
slow-roll approximation employed in the subsequent analysis.

\subsubsection{Second order action and linear perturbations}
\label{perturb}

We now turn to the study of linear cosmological perturbations and the
derivation of the second-order action governing the dynamics of primordial
fluctuations. This analysis allows us to determine the scalar and tensor power
spectra generated during inflation within scalar-torsion $f(T,\phi)$ gravity,
and to identify potential imprints of torsion and local Lorentz symmetry
breaking on cosmological observables.

In order to study primordial density fluctuations, we start from the
Arnowitt-Deser-Misner (ADM) decomposition of the tetrad field
\cite{Wu:2011kh}
\begin{eqnarray}
&&e_{~\mu }^{0}=\left( \mathcal{N},\mathbf{0}\right) ,\:\:\:\: e_{~\mu }^{a}=\left(
\mathcal{N}^{a},h_{~i}^{a}\right) ,  \label{ADM1} \\
&&e_{0}^{~\mu }=\left( 1/\mathcal{N},-\mathcal{N}^{i}/\mathcal{N}\right) ,\:\:\:\: e_{a}^{~\mu }=\left(
0,h_{a}^{~i}\right) ,  \label{ADM2}
\end{eqnarray}
where $\mathcal{N}$ denotes the lapse function and $\mathcal{N}^{i}=h_{a}^{~i}\mathcal{N}^{a}$ is the shift vector, with $h_{~j}^{a}h_{a}^{~i}=\delta _{j}^{i}$. The quantity $h_{~i}^{a}$ represents the induced tetrad field on spatial hypersurfaces. 

We work within the uniform field gauge, defined by $\delta\phi=0$, in which the
scalar-field perturbation is absorbed into the metric degrees of freedom. A
convenient parametrization of the perturbations is then given by
\begin{equation}
\mathcal{N}=1+\alpha ,\:\:\:\:\mathcal{N}^{a}=a^{-1}e^{-\mathcal{R}}\delta _{~i}^{a}\partial ^{i}{%
\psi },\:\:\:\: h_{~i}^{a}=ae^{\mathcal{R}}\delta _{~j}^{a}\delta _{~i}^{j},
\label{Uniform_Field_Gauge}
\end{equation}
which leads to the corresponding perturbed metric \cite{DeFelice:2011uc}
\begin{eqnarray}
ds^{2} &=&-\left[ \left( 1+\alpha \right) ^{2}-a^{-2}e^{-2\mathcal{R}}\left(
\partial \psi \right) ^{2}\right] dt^{2} 
+2\partial _{i}{\psi }dtdx^{i}+a^{2}e^{2\mathcal{R}}\delta
_{ij}dx^{i}dx^{j}.
\end{eqnarray}

In scalar-torsion gravity, the breaking of local Lorentz invariance gives rise
to additional degrees of freedom, which can be systematically incorporated as
Goldstone modes associated with the symmetry breaking
\cite{Bluhm:2004ep,Bluhm:2007bd}. These modes are introduced through a local
Lorentz rotation of the tetrad field. Under the transformation
\begin{equation}
\Lambda _{~B}^{A}=\left( e^{\chi }\right) _{~B}^{A}=\delta _{~B}^{A}+\chi
_{~B}^{A}+\frac{1}{2}\chi _{~C}^{A}\chi _{~B}^{C}+\mathcal{O}(\chi ^{3}),
\label{Lorentz_Transf}
\end{equation}
and keeping the spin connection fixed to its background value, the full tetrad
field can be written as
\begin{eqnarray}
e_{~\mu }^{\prime A} &=&\left( e^{\chi }\right) _{~B}^{A}e_{~\mu }^{B}, 
\notag \\
&=&e_{~\mu }^{A}+\chi _{~B}^{A}e_{~\mu }^{B}+\frac{1}{2}\chi _{~C}^{A}\chi
_{~B}^{C}e_{~\mu }^{B}+\mathcal{O}(\chi ^{3}) .
\label{Transf_tetrad}
\end{eqnarray}

The antisymmetric matrix $\chi_{AB}=-\chi_{BA}$ can be parametrized as
\begin{equation}
\chi _{~B}^{0}=\left( 0,\chi _{b}\right) ,\:\:\:\:
\chi _{~B}^{a}=\left( \chi ^{a},B_{~b}^{a}\right) ,
\end{equation}
where $\chi^{a}=\eta^{ab}\chi_{b}$ and $B_{ab}=-B_{ba}$. Defining the spatial
vector $\chi^{i}=h_{a}^{~i}\chi^{a}=\partial_{i}\beta+\chi_{i}^{(T)}$ and the
spatial antisymmetric tensor
$B_{ij}=h_{~i}^{a}h_{~j}^{b}B_{ab}=-B_{ji}=-\epsilon_{jik}B^{k}$, one identifies 
a
scalar mode $\beta$, a transverse vector mode $\chi_{i}^{(T)}$, and a
(pseudo-)vector mode $B_{i}$
\cite{Wu:2016dkt,Golovnev:2018wbh}.

Now, expanding the action up to second order in the curvature perturbation
$\mathcal{R}$, one obtains the quadratic action
\begin{equation}
S_{s}^{(2)} = \int dt\, d^{3}x\, a^3 Q_{s}
\left[
\dot{\mathcal{R}}^{2}
- \dfrac{c^{2}_{s}}{a^{2}} (\partial \mathcal{R})^2
- m^{2} \mathcal{R}^2
\right],
\label{SecOrderSM}
\end{equation}
where the coefficients are given by
\begin{eqnarray}
Q_s &=& \dfrac{3 w_{1} H^{2} + w_{3}}{H^{2}}
=\frac{PX}{H^2}, \label{Qs} \\ 
c^{2}_{s}&=& 1 , \label{cs} \\
m^{2}&=& \dfrac{\dot{w_{2}}}{w_{2}}
\left(
3 H + \dfrac{\dot{Q}_{s}}{Q_{s}}
- 2 \dfrac{\dot{w_{2}}}{w_{2}}
+ \dfrac{\ddot{w_{2}}}{\dot{w}_{2}}
+ \dfrac{w_{1}\dot{w_{2}}}{w_{6} Q_{s}}
\right) .
\label{m2}
\end{eqnarray}

Physical viability requires the absence of ghost and gradient instabilities, which translates into the conditions $Q_s>0$ and $c_s^2>0$. These requirements are satisfied by Eqs.~\eqref{Qs} (for $P>0$) and~\eqref{cs}. In terms of the slow-roll parameters introduced in the previous subsection,
these expressions simplify considerably. In particular, one finds
\begin{equation}
Q_{s}=w_{2}\,\delta_{PX},
\end{equation}
and it is convenient to define
\begin{equation}
\eta_{Q}
=\frac{\dot{Q}_{s}}{H Q_{s}}
=\delta_{P}+2\delta_{\phi}+2\epsilon .
\end{equation}
Similarly, the effective mass term can be expressed as
\begin{eqnarray}
\eta_{\mathcal{R}}
=\frac{m^{2}}{3H^{2}}
=\delta_{f_{,T}}
\left[
1+\left(1+\frac{\delta_{fX}}{\delta_{PX}}\right)
\frac{\delta_{f_{,T}}}{\delta_{f\dot{H}}}
\right] .
\label{mass_term}
\end{eqnarray}

The scalar power spectrum of curvature perturbations is then given by \cite{Gonzalez-Espinoza:2020azh}
\begin{eqnarray}
\mathcal{P}_{s}(k)
&\equiv &
\frac{k^{3}}{2\pi^{2}}
\left|\mathcal{R}_{k}(\tau)\right|^{2},
\notag \\
&\simeq &
\frac{H_{k}^{2}}{8\pi^{2}Q_{sk}}
\left[
1+2\eta_{\mathcal{R}}
\ln\left(\frac{k}{aH}\right)
\right] .
\label{Ps}
\end{eqnarray}
Additionally, the scalar spectral index is defined as
\begin{equation}
n_{s}-1
\equiv
\left.\frac{d\ln\mathcal{P}_{s}(k)}{d\ln k}\right|_{k=aH}
=-2\epsilon-\eta_{Q}+2\eta_{\mathcal{R}} ,
\label{ns_fTphi}
\end{equation}
which explicitly shows how the effects of local Lorentz violation enter the
scalar spectrum through the term $2\eta_{\mathcal{R}}$ at first order in the
slow-roll approximation. Finally, the running of the scalar spectral index is given by
\be
\alpha_{s}\equiv
\frac{dn_{s}}{d\ln k}
=
2\epsilon\eta
+\eta_{Q}\kappa_{Q}
-2\eta_{\mathcal{R}}\kappa_{\mathcal{R}} .
\ee
We mentioned that in the above expressions we have adopted the standard hierarchy of slow-roll parameters and we have
introduced
\be
\eta=\frac{\dot{\epsilon}}{H\epsilon},
\:\:\:\:
\kappa_{i}=\frac{\dot{\eta}_{i}}{H\eta_{i}},
\:\:\:
i=Q,\mathcal{R}.
\ee

We now turn to tensor perturbations, which encode the primordial gravitational
wave sector. Starting from the ADM decomposition of the tetrad field given in
 \eqref{ADM1} and \eqref{ADM2}, and working again in the uniform field gauge
$\delta\phi=0$, we choose the parametrization \cite{Wu:2011kh}
\begin{equation}
\mathcal{N}=1,\:\:\: \mathcal{N}^{a}=0, \:\:\: h^{a}_{~i}=a\!\left(\delta^{a}_{~i}
+\frac{1}{2}\gamma^{a}_{~i}\right).
\end{equation}
The corresponding induced three-metric then takes the form
\begin{eqnarray}
g_{ij}
=\eta_{ab} h^{a}_{~i} h^{b}_{~j}
=a^{2}\left[
\delta_{ij}
+h_{ij}
+\frac{1}{4}\gamma_{k i}\gamma^{k}_{~j}
\right],
\end{eqnarray}
where we have defined
\begin{equation}
h_{ij}
=\frac{1}{2}\eta_{ab}
\left(
\delta^{a}_{~i}\gamma^{b}_{~j}
+\delta^{b}_{~j}\gamma^{a}_{~i}
\right)
=\frac{1}{2}\left(\gamma_{ij}+\gamma_{ji}\right),
\end{equation}
and $\gamma^{a}_{~j}=\gamma^{i}_{~j}\delta^{a}_{~i}$.

Using the tetrad formalism, one can derive the second-order action for the
tensor modes. Decomposing the tensor perturbation as
$h_{ij}=h_{+}e^{+}_{ij}+h_{\times}e^{\times}_{ij}$, the quadratic action reads \cite{Weinberg:2008zzc}
\begin{equation}
S_{T}^{(2)}
=\sum_{\lambda}
\int dt\, d^{3}x\, a^{3} Q_{T}
\left[
\dot{h}_{\lambda}^{2}
-\frac{c_{T}^{2}}{a^{2}}
\left(\partial h_{\lambda}\right)^{2}
\right],
\label{Tensor_Modes}
\end{equation}
where $\lambda=+,\times$ denotes the two tensor polarization states. The
background-dependent coefficient $Q_{T}$ is given by 
\begin{equation}
Q_{T}= - \dfrac{1}{2} f_{,T},
\end{equation}
while the squared propagation speed of tensor modes is
\begin{equation}
c_{T}^{2}=1 .
\end{equation}
Finally, note that the absence of ghost instabilities in the tensor sector therefore requires
$f_{,T}<0$ \cite{Gonzalez-Espinoza:2020azh}.

The power spectrum of tensor perturbations is obtained as \cite{Gonzalez-Espinoza:2020azh}
\begin{equation}
\mathcal{P}_{T}
\equiv \sum_{\lambda}\frac{k^3}{2\pi^2}\left|h_{\lambda}(k)\right|^2\simeq \frac{H_{k}^{2}}{2\pi^{2}Q_{Tk}},
\end{equation}
where $H_{k}$ and $Q_{Tk}$ are evaluated at horizon crossing, $k=aH$. Hence, the
corresponding tensor spectral index is then
\begin{equation}
n_{T}
\equiv
\left.
\frac{d\ln\mathcal{P}_{T}}{d\ln k}
\right|_{k=aH}
=-2\epsilon-\delta_{f_{,T}} .
\label{nT}
\end{equation}

Lastly, the tensor-to-scalar ratio, evaluated at horizon crossing, is given by
\begin{equation}
r
=\frac{\mathcal{P}_{T}}{\mathcal{P}_{s}}
\simeq
16\,\delta_{PX}
=16\left(\epsilon-\delta_{f_{,T}}\right) .
\label{r}
\end{equation}
Thus, combining \eqref{nT} and \eqref{r}, one obtains the modified consistency
relation
\begin{equation}
r
=8\left(-n_{T}-3\delta_{f_{,T}}\right),
\label{r_2}
\end{equation}
which explicitly differs from the standard single-field inflationary result
due to the presence of scalar-torsion interactions.

\section{Inflationary dynamics in scalar-torsion \texorpdfstring{$f(T,\phi)$}{f(T,φ)} gravity}
\label{Inflationarydynamics}

In this section we investigate the inflationary dynamics arising in
scalar-torsion $f(T,\phi)$ gravity. Building on the background equations and
the perturbative framework developed in the previous section, we analyze the
inflationary regime under the slow-roll approximation and derive the
corresponding expressions for the main inflationary observables. We first
present a general slow-roll formulation valid for arbitrary scalar-torsion
models, and then we focus on specific dynamical regimes that allow for analytical
progress. This analysis provides the theoretical basis for the Higgs-inflation
scenario studied in the subsequent sections, as well as for the comparison of
the model predictions with current cosmological observations.

\subsection{Slow-roll formulation}
\label{Slow_roll}

During the inflationary phase, the background evolution can be treated within
the slow-roll approximation, under which the field equations simplify
considerably. In this regime, the first Friedmann  equation (\ref{Fr1}) and the Klein-Gordon equation (\ref{KGequation})
reduce to \cite{Gonzalez-Espinoza:2021qnv}
\begin{eqnarray}
\label{Fr1approx}
f(T,\phi) &=& 2T f_{,T}, \\
3P(\phi)H\dot{\phi} &=& f_{,\phi}.
\label{eqS}
\end{eqnarray}

Following Ref.~\cite{Gonzalez-Espinoza:2020azh,Gonzalez-Espinoza:2021qnv}, we adopt the ansatz
\begin{equation}
f(T,\phi)
=-\frac{M_{Pl}^{2}T}{2}
-G(T)F(\phi)
-V(\phi),
\end{equation}
with $M_{Pl}$ the Planck mass, 
which allows for a non-minimal scalar-torsion interaction encoded in the
functions $G(T)$ and $F(\phi)$, together with a scalar potential $V(\phi)$. Moreover, 
for simplicity, we restrict to the canonical kinetic coupling
$P(\phi)=1$. Under these considerations, the first Friedmann equation (\ref{Fr1approx})
becomes
\begin{equation}
\frac{M_{Pl}^{2}T}{2}
=
V(\phi)
-
G(T)F(\phi)\,
T\,\frac{\partial}{\partial T}
\ln\!\left(\frac{G^{2}}{T}\right).
\label{T_phi}
\end{equation}
Since $T=6H^2$, this equation can be written in the standard slow-roll form
\begin{equation}
3H^{2}=8\pi\mathcal{G}_{\mathrm{eff}}V,
\end{equation}
where we have defined the effective gravitational coupling  
\begin{equation}
\mathcal{G}_{\mathrm{eff}}(T,\phi)
=
\frac{\mathcal{G}}
{1+\frac{2C(T)F(\phi)G(T)}{M_{Pl}^{2}T}}.
\label{Geffdef}
\end{equation}
with  
\begin{equation}
C(T)
\equiv
T\frac{\partial}{\partial T}
\ln\!\left(\frac{G^{2}}{T}\right).
\end{equation}

On the other hand, the scalar-field equation \eqref{eqS} yields
\begin{equation}
\frac{\dot{\phi}}{M_{Pl}H}
\simeq
-\left[
\frac{2G F_{,\phi}}{M_{Pl}T}
+
\frac{2V_{,\phi}}{M_{Pl}T}
\right],
\end{equation}
where, in what follows, we assume $\dot{\phi}<0$.

Furthermore, in \cite{Gonzalez-Espinoza:2020azh,Gonzalez-Espinoza:2021qnv} the authors have derived the scalar and tensor
power spectra of primordial fluctuations within scalar-torsion
$f(T,\phi)$ gravity. In particular, the scalar power spectrum of curvature
perturbations is given by
\begin{eqnarray}
\mathcal{P}_{s}(k)
&\equiv&
\frac{k^{3}}{2\pi^{2}}
\left|\mathcal{R}_{k}(\tau)\right|^{2}
\simeq
\frac{H_{k}^{2}}{8\pi^{2}Q_{sk}}
\left[
1+2\eta_{\mathcal{R}}
\ln\!\left(\frac{k}{aH}\right)
\right],
\notag \\
&\simeq&
\frac{T}
{96\pi^{2}
\left[
\frac{GF_{,\phi}}{T}
+
\frac{V_{,\phi}}{T}
\right]^{2}} ,
\label{scalar_Power}
\end{eqnarray}
where all quantities are evaluated at horizon crossing, \emph{i.e.} at $k=aH$. The mass term associated with the parameter $\eta_{\mathcal{R}}$ becomes
{\footnotesize 
\bea
&& \eta_{\mathcal{R}}
=
\frac{m^{2}}{3H^{2}}
=
\delta_{f_{,T}}
\left[
1
+
\left(
1+\frac{\delta_{fX}}{\delta_{PX}}
\right)
\frac{\delta_{fX}}{\delta_{f\dot{H}}}
\right],\nonumber\\
&& =\frac{2}{M_{Pl}^2 T^2}\Bigg[-\frac{2 \left(G F_{,\phi}+V_{,\phi}\right)^2}{1+\frac{2 F G_{,T}}{M_{Pl}^2}}+\frac{2 \left(F_{,\phi} \left(G-2 T G_{,T}\right)+V_{,\phi}\right) \left(F_{,\phi} \left(G-T G_{,T}\right)+V_{,\phi}\right)}{1+\frac{2 F \left(2 T G_{,TT}+G_{,T}\right)}{M_{Pl}^2}}-\frac{M_{Pl}^2 T F_{,\phi}^2 G_{,T}^2}{F G_{,TT}}\Bigg].\nonumber\\
\eea}
Hence, the scalar spectral index $n_{s}$, corresponding to the spectrum
$\mathcal{P}_{s}(k)$, is defined as
$n_{s}-1\equiv d\ln\mathcal{P}_{s}(k)/d\ln k$,
leading to \cite{Gonzalez-Espinoza:2021qnv}:
{\footnotesize
\bea
n_{s}-1 &\equiv& \left.\frac{d \ln{\mathcal{P}_{s}(k)}}{d\ln{k}}\right|_{k=a H}=-4\epsilon-
\delta_{P}-2\delta_{\phi}+2 \eta_{\mathcal{R}},\nonumber\\
&\simeq& -\frac{4 V_{,\phi}^2 }{M_{Pl}^2 T^2}\left[\frac{2}{1+\frac{2 F G_{,T}}
{M_{Pl}^2}}+\frac{1}{1+\frac{2 F \left(2 T G_{,TT}+G_{,T}\right)}{M_{Pl}^2}}\right]- \frac{8 F_{,\phi} V_{,\phi}}{M_{Pl}^2 T^2} \left[\frac{2 G}{1+\frac{2 F G_{,T}}
{M_{Pl}^2}}-\frac{T G_{,T}-G}{1+\frac{2 F \left(2 T G_{,TT}+G_{,T}\right)}{M_{Pl}^2}}\right]\nonumber\\
&& -\frac{4 F_{,\phi}^2}{M_{Pl}^2 T^2} \Bigg[\frac{2 G^2}{1+\frac{2 F G_{,T}}
{M_{Pl}^2}}-\frac{G\left(2 T G_{,T}-G\right)}{1+\frac{2 F \left(2 T G_{,TT}+G_{,T}\right)}{M_{Pl}^2}} +\frac{ M_{Pl}^2 T G_{,T}^2}{F G_{,TT}}\Bigg]+\frac{4 F_{,\phi\phi} G}{T}+\frac{4 V_{,\phi\phi}}{ T}.
\label{nS1}
\eea}
Additionally, the running of the scalar spectral index becomes
\begin{equation}
    \alpha_s \equiv \frac{dn_s}{d\ln k} = \frac{\sqrt{6}}{T^{7/2}}
    \left[\Theta_1 + 4(\Theta_2 + \Theta_3)+ \Theta_4 \right],
\end{equation}
where $\Theta_1$, $\Theta_2$, $\Theta_3$ and $\Theta_4$ are defined in Appendix \ref{func_alpha} and can be evaluated straightforwardly once the slow-roll parameters are specified.

We now turn to the tensor sector. The power spectrum of tensor perturbations
is given by
\begin{equation}
\mathcal{P}_{T}
=
\frac{H_{k}^{2}}{2\pi^{2}Q_{Tk}},
\end{equation}
where all quantities are evaluated at horizon crossing, $k=aH$, and
\begin{equation}
Q_{Tk}
=
-\left.\frac{f_{,T}}{2}\right|_{k=aH}.
\end{equation}
The corresponding tensor-to-scalar ratio then takes the form
\cite{Gonzalez-Espinoza:2021qnv}
\bea
r
\equiv
\frac{\mathcal{P}_{T}}{\mathcal{P}_{s}}
\simeq
16\,\delta_{PX}
\simeq
\frac{
32\left[G F_{,\phi}+V_{,\phi}\right]^{2}
}{
M_{Pl}^{2}T^{2}
\left(
1+\frac{2F G_{,T}}{M_{Pl}^{2}}
\right)
}.
\label{r1}
\eea

Finally, in order to quantify the duration of inflation, we introduce the number of
e-folds $N$, which measures the logarithmic growth of the scale factor from a
given time $t$ up to the end of inflation at $t_{f}$, and  is defined as
\begin{equation}
N
\equiv
\int_{t}^{t_{f}} H\,dt
=
\int_{\phi}^{\phi_{f}}
\frac{H}{\dot{\phi}}\,d\phi .
\end{equation}
Using this relation, one can derive an analytical expression for the number of
e-folds as a function of the inflaton field, $N(\phi)$. Lastly, differentiating with
respect to $\phi$, one finds
\bea
\frac{dN}{d\phi}
=
\phi_{,N}^{-1}
=
\frac{1}{M_{Pl}}
\left[
\frac{2G F_{,\phi}}{T M_{Pl}}
+
\frac{2V_{,\phi}}{M_{Pl}T}
\right]^{-1}
>0,
\label{dNdphi}
\eea
where we have used $\dot{\phi}=-\phi_{,N}H$.

\subsection{Dominant-coupling (high-energy) regime}
\label{Strong_Coupling}

We now focus on the dominant-coupling, or high-energy, regime of the theory,
which is particularly relevant for the inflationary phase. This regime is
defined by the condition that the scalar-torsion interaction dominates over
the Einstein-Hilbert term, namely
\be
T \ll \left| G(T)F(\phi)C(T)\right|/M_{Pl}^{2}
\sim V/M_{Pl}^{2}.
\ee
In this limit, the effective gravitational dynamics is mainly governed by the
non-minimal scalar-torsion coupling.

We further assume a power-law form for the torsion-dependent function,
\begin{equation}
G(T)\sim T^{s},
\label{ansatzG}
\end{equation}
where $s$ is a positive constant. Under this assumption, the slow-roll first Friedmann equation
  \eqref{T_phi} simplifies significantly and reduces to
\be
\frac{M_{Pl}^{2}}{2}T+(2s-1)F\,T^{s}
\simeq
(2s-1)F\,T^{s}
=
V ,
\label{TEq1}
\ee
where, in the dominant-coupling regime, the contribution proportional to
$M_{Pl}^{2}T$ becomes subleading and can be neglected. 
Furthermore, in this approximation, the effective gravitational constant (\ref{Geffdef}) takes the simple form
\be
\mathcal{G}_{eff}
=
\frac{\mathcal{G}}{1+\beta^{-1}},
\ee
where we have defined the dimensionless parameter
\be
\beta
\equiv
\frac{M_{Pl}^{2}T^{1-s}}{2(2s-1)F}.
\label{beta_strong}
\ee
The dominant-coupling regime corresponds to $\mathcal{G}_{eff}/\mathcal{G}\ll
1$, or equivalently $\beta\ll 1$, indicating a strong suppression of the
effective gravitational interaction. Hence, solving Eq.~\eqref{TEq1} in this limit, one obtains
\be
T
\simeq
\left[
\frac{V}{(2s-1)F}
\right]^{1/s}.
\label{Strong_T}
\ee
We note that, assuming a positive coupling function $F(\phi)$, consistency of
this solution requires $s>1/2$.

Therefore, as we can see,  once the scalar potential $V(\phi)$ and the non-minimal coupling
function $F(\phi)$ are specified, the torsion scalar can be expressed as
$T=T(\phi)$. Furthermore, using Eq.~\eqref{dNdphi}, one can invert the
relation to obtain $\phi=\phi(N)$ and consequently $T=T(N)$. Although a
closed-form analytical solution is not always available, simple choices for
$V(\phi)$ and $F(\phi)$ often allow for analytic progress, while more general
cases can be treated numerically.
Lastly, within the high-energy approximation and for the ansatz \eqref{ansatzG}, the
scalar spectral index and the tensor-to-scalar ratio can be expressed
directly as functions of the number of e-folds $N$ 
\cite{Gonzalez-Espinoza:2021qnv}.
In particular, one finds
\bea
\label{ns_phi_Strong}
n_{s}(N)-1
&\simeq&
-\frac{2(2s-1)V_{,N}}{sV}
-\frac{(5s-2)F_{,N}}{s(2s-1)F}
-\frac{2s^{2}V F_{,N}^{2}}
{(s-1)(2s-1)F\left[VF_{,N}+(2s-1)FV_{,N}\right]}
\nonumber\\
&&
+\frac{(2s-1)F V_{,NN}}
{VF_{,N}+(2s-1)FV_{,N}}
+ 
\frac{V F_{,NN}}
{VF_{,N}+(2s-1)FV_{,N}},
\eea
{\small
\bea
&& \alpha_{s}(N)\simeq \frac{(5 s-2) F_{,NN}}{s (2 s-1) F}-\frac{2 s^2 V F_{,N}^3}{(s-1) (2 s-1) F^2 \left(V F_{,N}+2 s F V_{,N}-F V_{,N}\right)}-\frac{(2 s-1) F_{,N} V_{,NN}}{V F_{,N}+(2 s-1) F V_{,N}}-\nonumber\\
&& \frac{(2 s-1) F V^{(3)}}{V F_{,N}+(2 s-1) F V_{,N}}-\frac{F^{(3)} V}{V F_{,N}+(2 s-1) F V_{,N}}+\frac{2 s^2 F_{,N}^2 V_{,N}}{(s-1) (2 s-1) F \left(V F_{,N}+(2 s-1)F V_{,N}\right)}+\nonumber\\
&& \frac{4 s^2 V F_{,N} F_{,NN}}{(s-1) (2 s-1) F \left(V F_{,N}+(2 s-1) F V_{,N}\right)}-\frac{2 s^2 V F_{,N}^2 \left(V F_{,NN}+2 s F_{,N} V_{,N}+(2 s-1) F V_{,NN}\right)}{(s-1) (2 s-1) F \left(V F_{,N}+(2 s-1) F V_{,N}\right)^2}-\nonumber\\
&& \frac{F_{,NN} V_{,N}}{V F_{,N}+(2 s-1) F V_{,N}}+\frac{V F_{,NN} \left(V F_{,NN}+2 s F_{,N} V_{,N}+(2 s-1) F V_{,NN}\right)}{\left(V F_{,N}+(2 s-1) F V_{,N}\right)^2}-\frac{(5 s-2) F_{,N}^2}{s (2 s-1) F^2}+\nonumber\\
&& \frac{(2 s-1) F V_{,NN} \left(V F_{,NN}+2 s F_{,N} V_{,N}+(2 s-1) F V_{,NN}\right)}{\left(V F_{,N}+(2 s-1) F V_{,N}\right)^2}+\frac{2 (2 s-1) V_{,NN}}{s V}-\frac{2 (2 s-1) V_{,N}^2}{s V^2},
\label{alpha_strong}
\eea}
and
\bea
r(N)
\simeq
\frac{8F_{,N}}{sF}
+
\frac{8(2s-1)V_{,N}}{sV},
\label{r_phi_Strong}
\eea
respectively. 

\section{Higgs-like Inflation: Analytical Results}
\label{AnalyticalResults}

In this section we specialize the general formalism developed above to a
Higgs-like inflationary scenario within scalar-torsion $f(T,\phi)$ gravity.
By Higgs-like we mean that the inflaton is a generic scalar degree of
freedom whose potential has the same symmetry-breaking quartic structure as
the Higgs potential, without being identified with the Standard Model Higgs
field. Thus, the parameters of the scalar potential should be interpreted as
effective parameters of the inflationary sector, rather than as Standard
Model Higgs-sector quantities. By adopting specific functional forms for the
scalar potential and the non-minimal scalar-torsion coupling, we derive
analytical expressions for the inflationary observables and confront the
model predictions with current cosmological data.

We consider the Higgs-like symmetry-breaking potential
\be
V(\phi)=\frac{\lambda}{4}(\phi^{2}-\nu^{2})^{2},
\label{Higgs}
\ee
where $\nu$ denotes the symmetry-breaking scale of the effective scalar
sector, and $\lambda$ is an effective quartic self-coupling, fixed by the
amplitude of scalar perturbations. Since $\phi$ is not identified with the
Standard Model Higgs field, no Standard Model renormalization-group
constraints, electroweak-scale boundary conditions, or Higgs-sector
unitarity bounds are imposed in the present analysis. In addition, we assume
a power-law form for the non-minimal scalar-torsion coupling function,
namely
\be
F(\phi)=\xi\,\phi^{c},
\label{Coupling_F}
\ee
with $\lambda$, $\xi$, and $c$ being positive constants.
Using the dominant-coupling expression \eqref{Strong_T}, the torsion scalar
can be written explicitly as a function of the scalar field as
\be
T
=
\left[
\frac{\lambda}{4(2s-1)\xi}
\right]^{\frac{1}{s}}
\phi^{\frac{4-c}{s}} .
\label{Strong_T_p}
\ee
Substituting this result into \eqref{dNdphi} and assuming the large-field
regime $\phi^{2}\gg\nu^{2}$, we can integrate the field evolution to obtain
an explicit relation between the scalar field and the number of e-folds, namely
\be
\phi(N)
=
A\left(N-\tilde{N}\right)^{\tilde{d}},
\label{phi_N}
\ee
where
\be
A
=
\left[
\frac{2\lambda
\left(
\frac{c}{4(2s-1)}+1
\right)}
{\tilde{d}\left[\frac{\lambda}{4(2s-1)\xi}\right]^{\frac{1}{s}}}
\right]^{\tilde{d}},
\ee
and
\be
\tilde{N}
=
-\left(\frac{\phi_{f}}{A}\right)^{\frac{1}{\tilde{d}}},
\ee
with
\be
\tilde{d}=\frac{1}{\frac{4-c}{s}-2}.
\ee
Using this solution together with \eqref{scalar_Power}, the scalar power
spectrum evaluated at horizon crossing, $N=N_{*}$, can be written as
\be
\mathcal{P}_{s}(N_{*})
=
\frac{\lambda (4-c-8s)
\left[
c-4+2\left(c+2(s-2)\right)N_{*}
\right]^{3}}
{384\pi^{2}s^{3}(2s-1)} .
\ee
Thus, solving for the self-coupling parameter $\lambda$, we find
\be
\lambda(N_{*})
=
\frac{384\pi^{2}s^{3}(2s-1)\mathcal{P}_{s}}
{(4-c-8s)
\left[
c-4+2\left(c+2(s-2)\right)N_{*}
\right]^{3}} ,
\ee
while we recall that according to the latest Planck results the amplitude of primordial scalar
perturbations is $\mathcal{P}_{s}=2.141\times10^{-9}$ at the pivot scale
$k_{*}=0.05\,\mathrm{Mpc}^{-1}$ \cite{Planck:2018jri}.

In a similar manner, combining  \eqref{Strong_T_p} and \eqref{phi_N}, we
obtain  the non-minimal coupling parameter as
\be
\xi(N_{*})
=
\left[
\frac{8(2s-1)}{\tilde{d}}
\left(
\frac{c}{4(2s-1)}+1
\right)
\right]^{s\left(\frac{1}{2\tilde{d}}-1\right)}
\left[
\frac{\lambda}{4(2s-1)}
\right]^{1-s}
\left[
\frac{2(2s-1)}{M_{Pl}^{2}}
\right]^{-\frac{s}{2\tilde{d}}}
\beta^{-\frac{s}{2\tilde{d}}}
N_{*}^{s\left(\frac{1}{2\tilde{d}}-1\right)},
\ee
where $\beta$ is defined in \eqref{beta_strong}. Moreover, substituting  \eqref{Higgs}, \eqref{Coupling_F}, and \eqref{phi_N} into the
general expressions \eqref{ns_phi_Strong}, \eqref{alpha_strong}, and \eqref{r_phi_Strong}, we obtain
the scalar spectral index, its running, and the tensor-to-scalar ratio as
\bea
n_{s}(N_{*})-1
&=&
-\frac{
\frac{2c^{2}\tilde{d}}{(c+4)(s-1)}
-\frac{8(c-4)c\tilde{d}}{(c+4)(c+8s-4)}
+\frac{2(c-4)\tilde{d}}{s}
+12\tilde{d}+1
}
{N_{*}-\tilde{N}}
\nonumber\\
&\approx&
-\frac{
\frac{2c^{2}\tilde{d}}{(c+4)(s-1)}
-\frac{8(c-4)c\tilde{d}}{(c+4)(c+8s-4)}
+\frac{2(c-4)\tilde{d}}{s}
+12\tilde{d}+1
}
{N_{*}},
\label{ns_N}
\eea
\bea
\alpha_{s}(N_{*})=-\frac{\frac{2 c^2\tilde{d}}{(c+4) (s-1)}-\frac{8 (c-4) c \tilde{d}}{(c+4) (c+8 s-4)}+\frac{2 (c-4) \tilde{d}}{s}+12 \tilde{d}+1}{N_{*}^2},
\eea
and
\bea
r(N_{*})
=
\frac{8\tilde{d}(c+8s-4)}{s(N_{*}-\tilde{N})}
\approx
\frac{8\tilde{d}(c+8s-4)}{sN_{*}},
\label{r_N}
\eea
where in the last step we have used $N_{*}\gg\tilde{N}$, as typically expected
for horizon crossing.
Hence, combining \eqref{ns_N} and \eqref{r_N}, we obtain a direct relation
between the tensor-to-scalar ratio and the scalar spectral index, namely
\be
r(n_{s})
=
\frac{
8(1-n_{s})(s-1)(c+8s-4)^{2}
}
{
2(13c-76)s^{2}
+(c-4)(3c-22)s
-(c-4)^{2}
+80s^{3}
}.
\label{r_ns}
\ee
We recall that the latest cosmological constraints from the combined analysis of Planck, ACT, and DESI DR1 (P--ACT--LB) yield \cite{AtacamaCosmologyTelescope:2025blo}
\be
n_{s}=0.9743\pm 0.0034.
\label{n_s_C}
\ee
In addition, recent
BICEP/Keck XIII results place an upper bound on the tensor-to-scalar ratio, namely
\be
r<0.036,
\ee
at $95\%$ confidence level \cite{BICEPKeck:2021gln}.

We adopt the range $50\leq N_*\leq 60$, which is commonly used in
single-field slow-roll inflation. In the present scalar-torsion framework,
this range is used as a benchmark interval. A precise determination of
$N_*$ requires matching the comoving pivot scale at horizon exit,
$k_*=a_*H_*$, to the present scale, and therefore depends on the
post-inflationary expansion history. In minimally coupled Einstein gravity,
this matching is controlled by the inflationary energy scale, the energy
density at the end of inflation, the reheating energy density, and the
effective reheating equation of state. In the model considered here, the
matching is modified by the scalar-torsion sector, since the slow-roll
Friedmann equation contains the effective gravitational coupling
$\mathcal{G}_{\rm eff}(T,\phi)$ and the nonminimal interaction $F(\phi)G(T)$, with
$F(\phi)=\xi\phi^c$ and $G(T)\sim T^s$. Hence, the relation between
$H_*$, the scalar potential, and the inflationary energy scale is
model-dependent. A complete computation of $N_*$ would require the post-inflationary
evolution of the scalar-torsion system, including the end of slow roll,
reheating, the effective reheating equation of state, and the evolution of
$\mathcal{G}_{\rm eff}(T,\phi)$. This analysis is beyond the scope of the present
work. We therefore keep $N_*$ in the interval $50\leq N_*\leq 60$ and show
explicitly the dependence of the observables on $N_*$ in the analytical
expressions and numerical trajectories.

Using these observational
constraints together with \eqref{r_ns}, one can derive the allowed regions for
the parameters $c$ and $s$. Since the horizon crossing typically occurs for
$N_{*}\simeq50$ - $60$, the inflationary predictions depend primarily on the
parameters $c$, $s$, and $N_{*}$.

\begin{figure}[!th]
    \centering
    \begin{subfigure}[b]{0.49\textwidth}
        \centering
        \includegraphics[width=\textwidth]{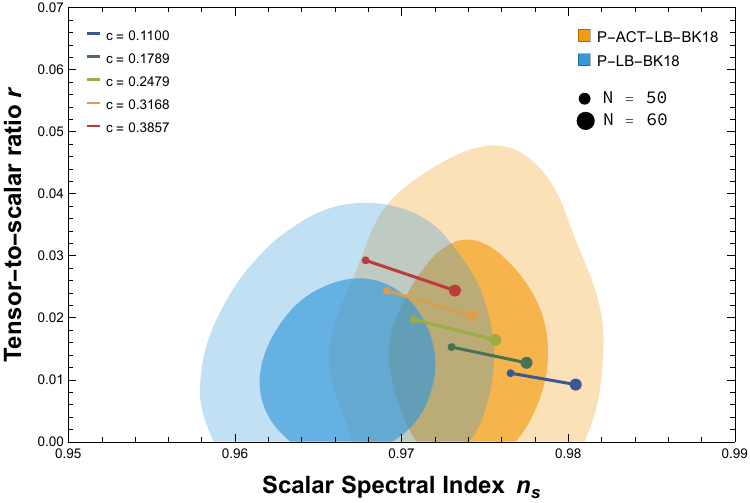}
        \caption{Fixed $s$, varying $c$}
        \label{fig:ns_r_fixed_s}
    \end{subfigure}
    \hfill 
    \begin{subfigure}[b]{0.49\textwidth}
        \centering
        \includegraphics[width=\textwidth]{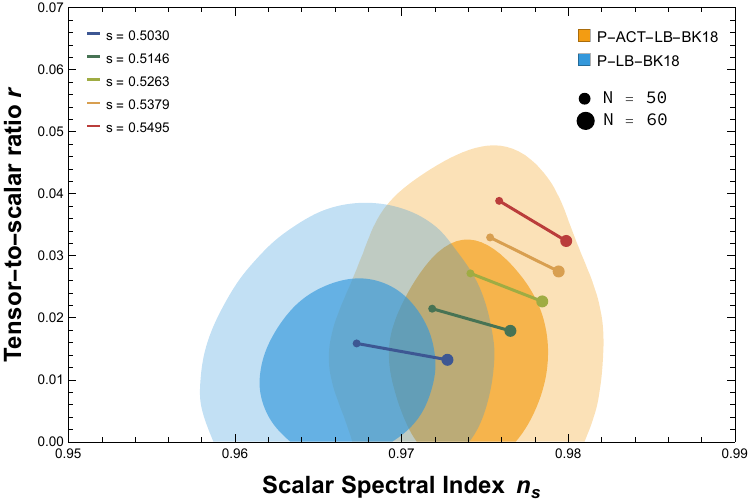}
        \caption{Fixed $c$, varying $s$}
        \label{fig_he:ns_r_vary_s}
    \end{subfigure}
    
    \caption{Predictions of the Higgs-like inflation model in the ($n_s - r$) plane evaluated at the pivot scale $k_* = 0.05 Mpc^{-1}$. \textbf{(a)} Predictions for $c \in [0.1100, 0.3857]$ and fixed $s = 0.5110$. \textbf{(b)} Predictions for varying $s \in [0.5030,0.5495]$ and fixed $c = 0.2479$. The endpoints of each trajectory correspond to $N = 50$ (small circle) and $N = 60$ (large circle). The blue shaded regions represent the $68\%$ and $95\%$ confidence contours from Planck 2018+Lensing+BICEP/Keck 2021, while the orange shaded regions show the corresponding constraints from the combined Planck 2018+ACT DR6+Lensing+BICEP/Keck 2021 analysis.\cite{Planck:2018jri,AtacamaCosmologyTelescope:2025blo,BICEPKeck:2021gln}}
    \label{fig_he:combined_ns_r}
\end{figure}

\begin{figure}[!t]
    \centering
    \begin{subfigure}[b]{0.40\textwidth}
        \centering
        \includegraphics[width=\textwidth]{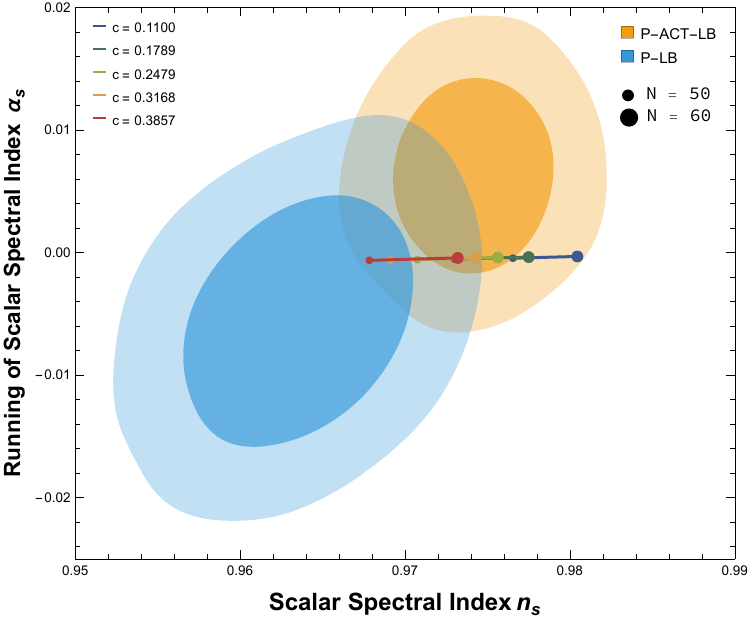}
        \caption{Fixed $s$, varying $c$}
        \label{fig:ns_r_fixed_s2}
    \end{subfigure}
    \hspace{1cm} 
    \begin{subfigure}[b]{0.40\textwidth}
        \centering
        \includegraphics[width=\textwidth]{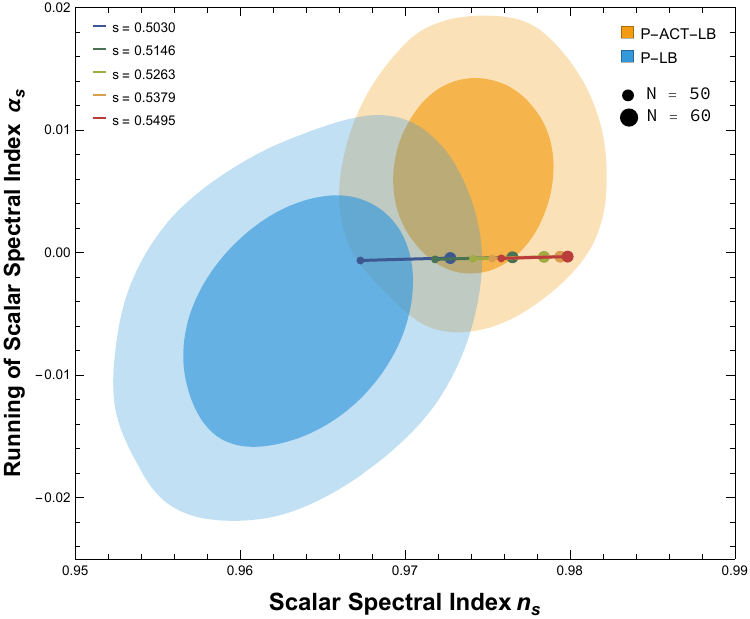}
        \caption{Fixed $c$, varying $s$}
        \label{fig:ns_r_vary_s}
    \end{subfigure}
    \caption{Predictions of the Higgs-like inflation model in the ($n_s-\alpha_s$) plane evaluated at the pivot scale $k_* = 0.05\,\mathrm{Mpc}^{-1}$. \textbf{(a)} Predictions for $c \in [0.1100,0.3857]$ and fixed $s = 0.5110$. \textbf{(b)} Predictions for varying $s \in [0.5030,0.5495]$ and fixed $c = 0.2479$. The endpoints of each trajectory correspond to $N = 50$ (small circle) and $N = 60$ (large circle). The blue shaded regions represent the $68\%$ and $95\%$ confidence contours from the Planck 2018+Lensing+BAO analysis, while the orange shaded regions show the corresponding constraints from the combined Planck 2018+ACT DR6+DESI BAO analysis \cite{Planck:2018jri,AtacamaCosmologyTelescope:2025blo}. Theoretical trajectories illustrate the evolution of the scalar spectral index $n_s$ and its running $\alpha_s$ over the range $N \in [50,60]$.} 
    \label{fig_he:alpha_vs_ns}
\end{figure}

\begin{table*}[t]
\centering
\caption{Model predictions for varying $c$ (Top Block) and varying $s$ (Bottom Block). Observables are calculated at $N=50$ and $N=60$. Derived parameters are evaluated at $N=60$(horizon crossing).}
\label{tab_he:model_results}
\renewcommand{\arraystretch}{1.3} 
\resizebox{\textwidth}{!}{%
\begin{tabular}{!{\vrule width 1.5pt}c|c!{\vrule width 1.5pt}c|c|c!{\vrule width 1.5pt}c|c|c!{\vrule width 1.5pt}c|c|c|c|c|c!{\vrule width 1.5pt}}
\noalign{\hrule height 1.5pt}

\multicolumn{2}{!{\vrule width 1.5pt}c!{\vrule width 1.5pt}}{\textbf{Input}} & 
\multicolumn{3}{c!{\vrule width 1.5pt}}{\textbf{Observables ($N=50$)}} & 
\multicolumn{3}{c!{\vrule width 1.5pt}}{\textbf{Observables ($N=60$)}} & 
\multicolumn{6}{c!{\vrule width 1.5pt}}{\textbf{Derived Parameters ($N=60$)}} \\
\noalign{\hrule height 1.0pt} 
$c$ & $s$ & $n_s$ & $r$ & $\alpha_s$ & $n_s$ & $r$ & $\alpha_s$ & $\xi$ & $\lambda$ & $\epsilon$ & $\eta$ & $H$ & $E_{\text{inf}}$ \\

\noalign{\hrule height 1.5pt}
0.1100 & 0.5110 & 0.9765 & 0.0110 & -4.69e-04 & 0.9765 & 0.0092 & -3.26e-04 & 0.2066 & 2.85e-15 & 0.0112 & 1.51e-02 & 2.35e-04 & 0.0720 \\
0.1789 & 0.5110 & 0.9730 & 0.0153 & -5.40e-04 & 0.9730 & 0.0127 & -3.75e-04 & 0.1603 & 2.28e-15 & 0.0112 & 1.36e-02 & 2.77e-04 & 0.0781 \\
0.2479 & 0.5110 & 0.9708 & 0.0197 & -5.85e-04 & 0.9708 & 0.0164 & -4.06e-04 & 0.1190 & 1.95e-15 & 0.0113 & 1.27e-02 & 3.14e-04 & 0.0832 \\
0.3168 & 0.5110 & 0.9691 & 0.0243 & -6.17e-04 & 0.9691 & 0.0203 & -4.29e-04 & 0.0859 & 1.75e-15 & 0.0114 & 1.20e-02 & 3.49e-04 & 0.0878 \\
0.3857 & 0.5110 & 0.9678 & 0.0292 & -6.43e-04 & 0.9678 & 0.0244 & -4.47e-04 & 0.0609 & 1.61e-15 & 0.0115 & 1.15e-02 & 3.83e-04 & 0.0919 \\

\noalign{\hrule height 1.5pt}
0.2479 & 0.5030 & 0.9708 & 0.0197 & -5.85e-04 & 0.9708 & 0.0164 & -4.06e-04 & 0.1190 & 1.95e-15 & 0.0113 & 1.27e-02 & 3.14e-04 & 0.0832 \\
0.2479 & 0.5146 & 0.9732 & 0.0244 & -5.36e-04 & 0.9732 & 0.0203 & -3.72e-04 & 0.0637 & 3.21e-15 & 0.0114 & 1.37e-02 & 2.58e-04 & 0.0754 \\
0.2479 & 0.5263 & 0.9746 & 0.0291 & -5.07e-04 & 0.9746 & 0.0243 & -3.52e-04 & 0.0472 & 4.28e-15 & 0.0115 & 1.43e-02 & 2.35e-04 & 0.0720 \\
0.2479 & 0.5379 & 0.9754 & 0.0339 & -4.91e-04 & 0.9754 & 0.0283 & -3.41e-04 & 0.0402 & 5.26e-15 & 0.0116 & 1.46e-02 & 2.23e-04 & 0.0701 \\
0.2479 & 0.5495 & 0.9759 & 0.0388 & -4.83e-04 & 0.9759 & 0.0324 & -3.35e-04 & 0.0368 & 6.19e-15 & 0.0116 & 1.48e-02 & 2.16e-04 & 0.0690 \\

\noalign{\hrule height 1.5pt}
\end{tabular}%
}
\end{table*}

In Fig.~\ref{fig_he:combined_ns_r}, we present the parametric plot of $n_{s}$ versus $r$ for
selected values of $c$ and $s$, and for $50\leq N_{*}\leq60$. The figure shows
that the model predictions lie well within the $1\sigma$ and $2\sigma$ confidence regions
of the combined data of ACT-2025, Planck-2018 and BICEP/Keck 2021(P-ACT-LB-BK) data.
In Fig. \ref{fig_he:alpha_vs_ns}, we present the marginalized constraints on the scalar spectral index $n_s$ and its running $\alpha_s$ at the 68\% and 95\% confidence levels. The blue contours correspond to the constraints obtained from the \textit{Planck} 2018 data, including post-lensing and baryon acoustic oscillation (BAO) measurements, while the orange contours represent the joint constraints derived from the combined analysis of Planck 2018, ACT DR6, and DESI BAO datasets.

\begin{figure}
    \centering
    \includegraphics[width=0.85\textwidth, height=0.60\textheight]{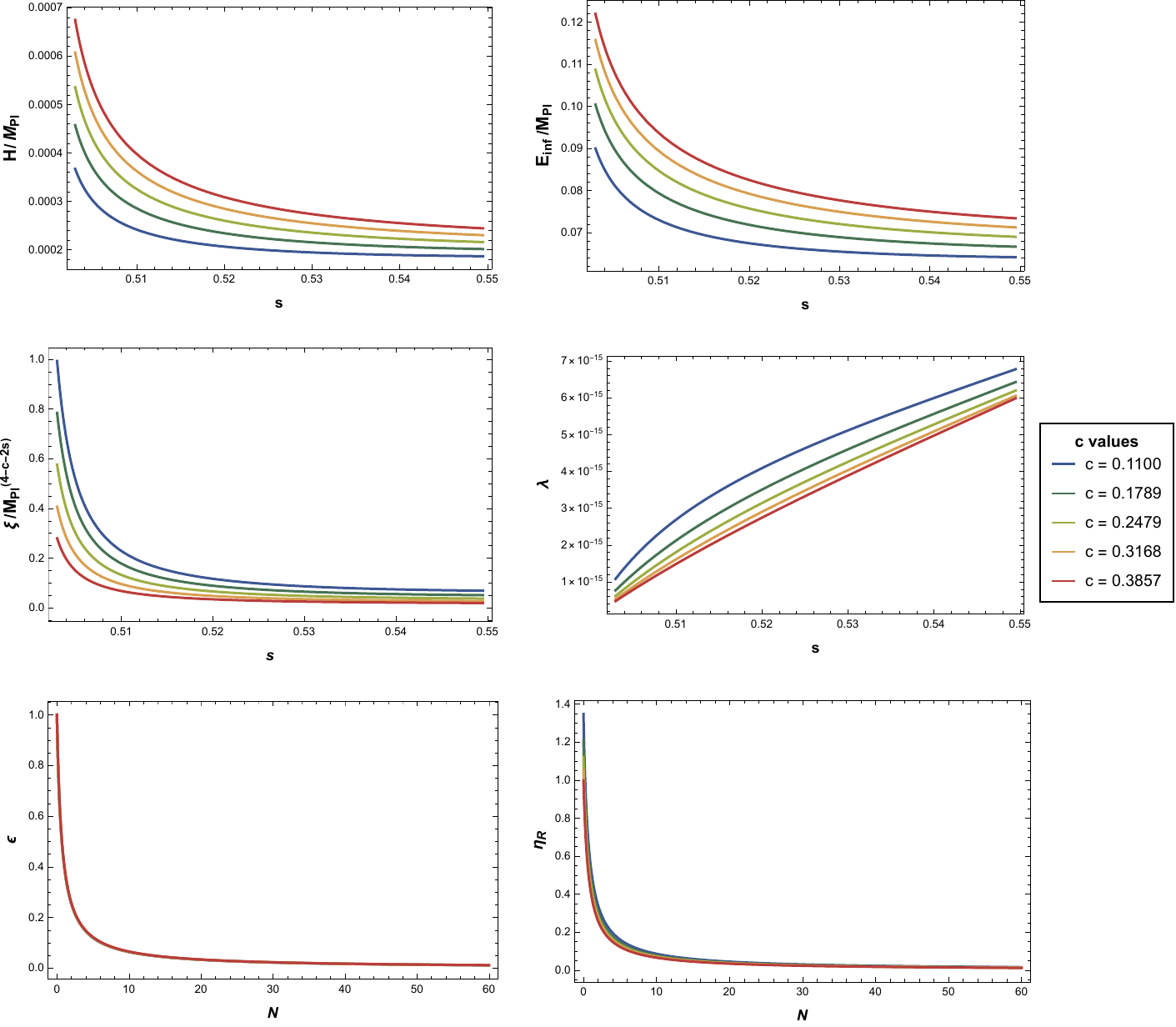}
    \caption{Background inflationary quantities in the Higgs-like inflation 
model with $f(T,\phi)$ gravity, obtained analytically within the dominant-coupling regime. \textit{Upper panels}: the Hubble parameter $H(N_\ast)/M_{\rm Pl}$, the inflationary energy scale $E_{\rm inf}/M_{\rm Pl}$, the non-minimal coupling $\xi(N_\ast)/M_{\rm Pl}^{4-c-2s}$, and the effective self-coupling $\lambda(N_\ast)$, all evaluated at horizon crossing $N_\ast = 60$ and displayed as functions of $s$ for $s \in [0.5030,\,0.5495]$. \textit{Lower panels}: the slow-roll parameters $\epsilon(N)$ and $\eta_R(N)$ as functions of the number of e-folds $N$, computed at the 
fixed representative value $s = 0.515$ over the full inflationary interval $N \in [60,\,0]$. Each curve corresponds to a distinct value of $c \in [0.1100,\,0.3857]$, illustrating how the torsion--scalar coupling modifies the inflationary dynamics across the explored 
parameter range.}
    \label{fig:all_parameter_plots}
\end{figure}

In Fig.~\ref{fig:all_parameter_plots}, we illustrate the behavior of
$H(N_{})$, $\epsilon(N_{})$, $\lambda(N_{})$, and $\xi(N_{})$,
evaluated at $N_{*}=60$, for $s\in[0.5030,0.5495]$ and for several
representative values of $c\in[0.1100, 0.3857]$. The figure shows that the Hubble scale, 
the inflationary energy scale, and the effective couplings vary smoothly 
with the torsion–scalar parameters, while the slow-roll parameters $\epsilon$ and
$\eta_{\mathcal R}$ remain well below unity over the entire inflationary interval. This confirms that the Higgs-like inflationary regime is dynamically stable and that the analytical reconstruction of the model parameters is consistent with sustained slow-roll inflation. Moreover, Table \ref{tab_he:model_results} summarizes the main results from Figs. \ref{fig_he:combined_ns_r}-\ref{fig:all_parameter_plots}. It lists predicted values of $n_s$, $r$, and $\alpha_s$ for representative $c$ and $s$ at $N=50$ and $N=60$, with derived quantities at $N=60$. The confidence regions align with current observational bounds and concisely summarize the information from the contour plots.

We close this section with a comment on the issue of perturbative unitarity. Unitarity
has played an important role in the discussion of Higgs inflation in curvature-based
gravitational frameworks. In the original Higgs inflation scenario, where the
Standard Model Higgs field is nonminimally coupled to the Ricci scalar, 
the theory exhibits a unitarity cutoff
well below the Planck scale for large values of the coupling, raising
concerns about the validity of the effective field theory during the inflationary
regime \cite{Burgess:2009ea,Barbon:2009ya}. This issue was later addressed in the
so-called ``New Higgs inflation'' scenario, where a nonminimal derivative
coupling between the Higgs field and the Einstein tensor was introduced, leading
to a restoration of perturbative unitarity up to the inflationary scale
\cite{Germani:2010gm,Germani:2011ua}. In the present work, however, 
the situation is fundamentally different. The
scalar field $\phi$ driving inflation is not identified with the Standard Model
Higgs field, but instead represents a generic Higgs-like inflaton characterized
by a quartic symmetry-breaking potential. As a result, the parameters of the potential are not constrained by Standard Model physics. Moreover, the
nonminimal interaction is not of the curvature-based form $\xi\phi^2R$, but is encoded in the scalar-torsion sector through $F(\phi)G(T)$, with
$F(\phi)=\xi\phi^c$ and $G(T)\sim T^s$. Therefore, the usual cutoff estimates
associated with Standard Model Higgs inflation cannot be directly applied to the present setup. Nevertheless, this does not by itself establish the absence of cutoff or
strong-coupling effects in the scalar-torsion effective theory. The actual
perturbative cutoff should instead be determined by expanding the
scalar-torsion action around the relevant inflationary background and
identifying the higher-dimensional operators that control scattering
amplitudes \cite{Hu:2023juh,Hu:2023xcf}. Such an analysis is beyond the scope of the present work.

\section{Higgs-like Inflation: Numerical analysis beyond the high-energy 
approximation}
\label{beyondhighenergy}

In this section we perform a numerical analysis of the Higgs-like inflationary scenario by integrating the slow-roll background equations without invoking the dominant-coupling, or high-energy, approximation employed in the previous section. Thus, the analysis goes beyond the high-energy approximation, but it remains within the slow-roll regime. In particular, we do not solve the exact background equations without slow-roll assumptions. The purpose of this numerical treatment is to test the robustness of the analytical high-energy results and to explore the inflationary dynamics in a more general slow-roll regime.

\begin{figure}[ht]
    \centering
    \begin{subfigure}[b]{0.32\textwidth}
        \centering
        \includegraphics[width=\linewidth, height=0.17\textheight]{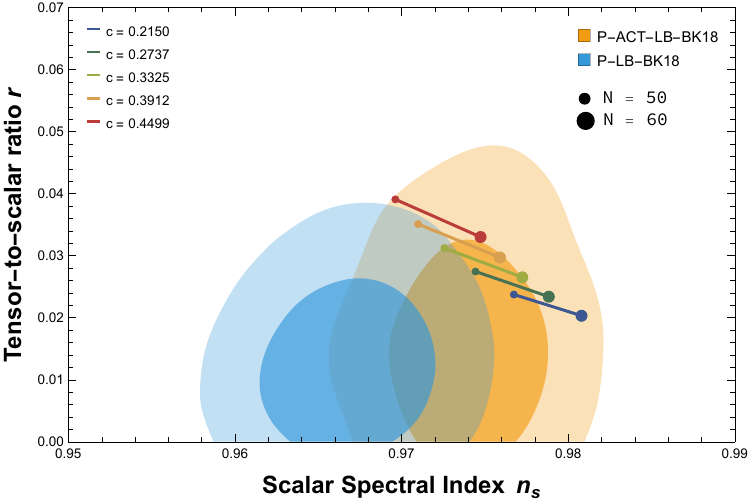}
        \caption{Fixed $s$ and $\gamma$, varying $c$ }
        \label{fig:ns_r_fixed_s_2_numerical}
    \end{subfigure}
    \hfill 
    \begin{subfigure}[b]{0.32\textwidth}
        \centering
        \includegraphics[width=\linewidth, height=0.17\textheight]{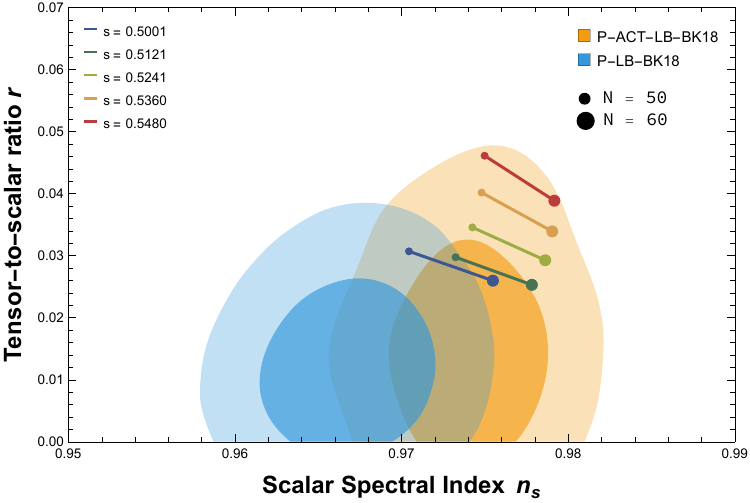}
        \caption{Fixed $c$ and $\gamma$, varying $s$}
        \label{fig:ns_r_vary_s_numerical}
    \end{subfigure}
    \hfill 
    \begin{subfigure}[b]{0.32\textwidth}
        \centering
        \includegraphics[width=\linewidth, height=0.17\textheight]{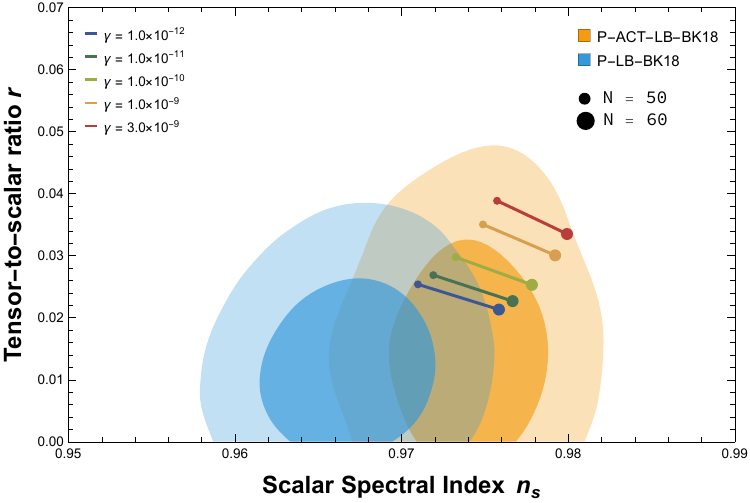}
        \caption{Fixed $c$ and $s$, varying $\gamma$}
        \label{fig:ns_r_vary_L0}
    \end{subfigure}
    
    \caption{Constraints in the $n_s-r$ plane comparing theoretical predictions with current CMB observations at pivot scale $k=0.05Mpc^{-1}$. 
    The blue shaded regions represent the $68\%$ and $95\%$ confidence contours from Planck 2018+Lensing+BICEP/Keck 2021, while the orange shaded regions show the corresponding constraints from the combined Planck 2018+ACT DR6+Lensing+BICEP/Keck 2021 analysis.\cite{Planck:2018jri}\cite{BICEPKeck:2021gln}\cite{AtacamaCosmologyTelescope:2025blo} 
    (a) Predictions for $s=0.5121$ and $\gamma=10^{-10}$, with the parameter $c$ varied in the range 
    $c \in [0.2150,\,0.4499]$. 
    (b) Predictions for fixed $c=0.3099$ and $\gamma=10^{-10}$, with $s$ varied in the range 
    $s \in [0.5001,\,0.5480]$. 
    (c) Predictions for fixed $c=0.3099$ and $s=0.5121$, with $\gamma$ varied over the indicated values. 
    In all panels, the endpoints of each theoretical trajectory correspond to the number of e-folds 
    $N = 50$ (small markers) and $N = 60$ (large markers).
    }
    \label{fig:combined_ns_r_numerical}
\end{figure}
\begin{figure}[ht]
    \centering
    
    \begin{subfigure}[b]{0.32\textwidth}
        \centering
        \includegraphics[width=\linewidth, height=0.19\textheight, keepaspectratio]{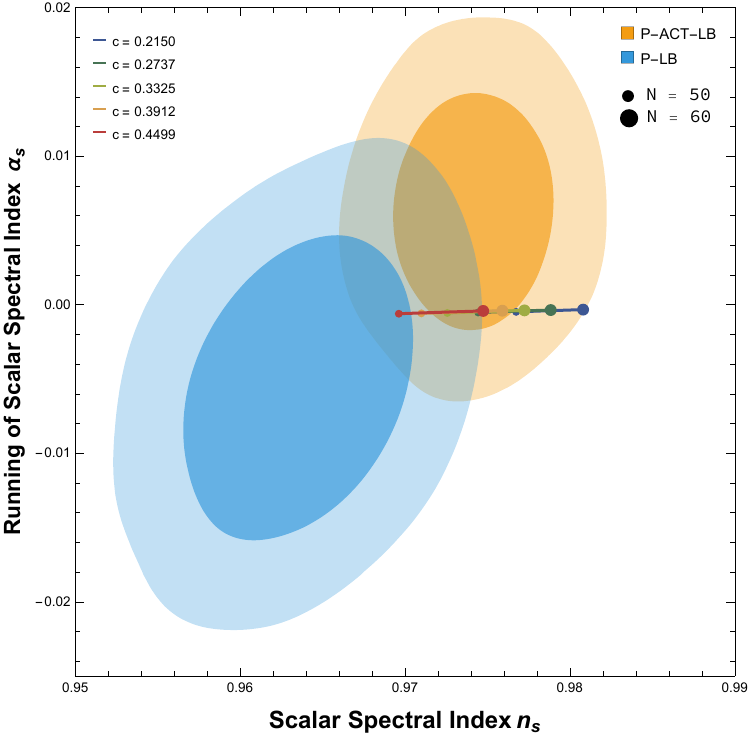}
        \caption{Fixed $s$ and $\gamma$, varying $c$}
        \label{fig:as_ns_fixed_s}
    \end{subfigure}
    \hfill 
    \begin{subfigure}[b]{0.32\textwidth}
        \centering
        \includegraphics[width=\linewidth, height=0.19\textheight, keepaspectratio]{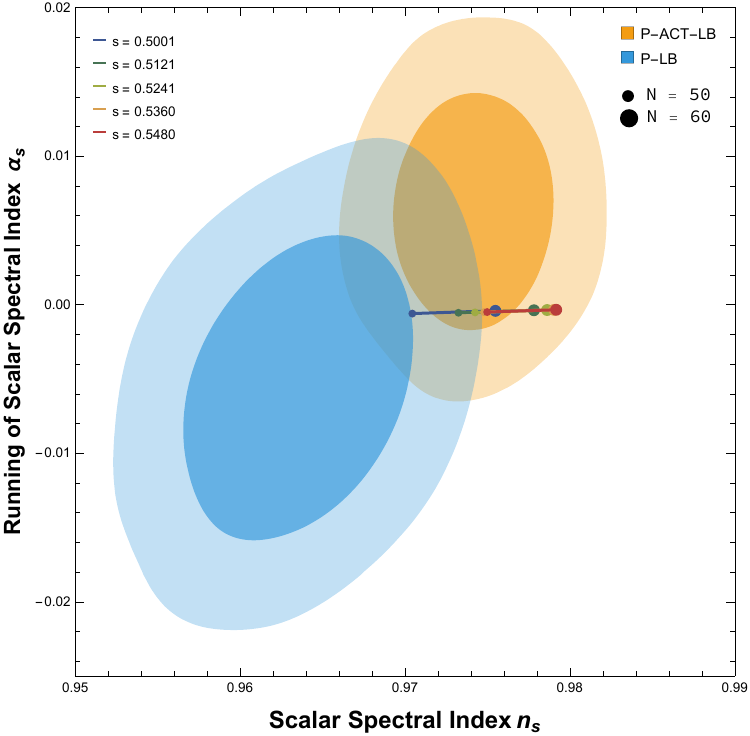}
        \caption{Fixed $c$ and $\gamma$, varying $s$}
        \label{fig:as_ns_vary_s}
    \end{subfigure}
    \hfill 
    \begin{subfigure}[b]{0.32\textwidth}
        \centering
        \includegraphics[width=\linewidth, height=0.19\textheight, keepaspectratio]{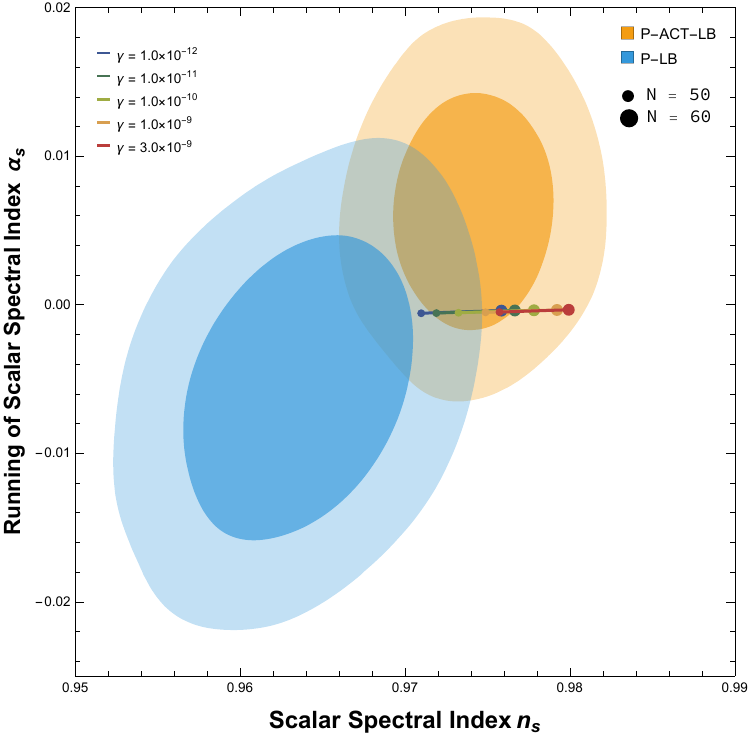}
        \caption{Fixed $c$ and $s$, varying $\gamma$}
        \label{fig:as_ns_vary_L0}
    \end{subfigure}
        \caption{Constraints in the $n_s-\alpha_s$ plane comparing theoretical predictions with current CMB observations at pivot scale $k=0.05Mpc^{-1}$. 
    The blue shaded regions represent the $68\%$ and $95\%$ confidence contours from the Planck 2018+Lensing+BAO analysis, while the orange shaded regions show the corresponding constraints from the combined Planck 2018+ACT DR6+DESI BAO analysis.
    (a) Predictions for $s=0.5121$ and $\gamma=10^{-10}$, with the parameter $c$ varied in the range 
    $c \in [0.2150,\,0.4499]$. 
    (b) Predictions for fixed $c=0.3099$ and $\gamma=10^{-10}$, with $s$ varied in the range 
    $s \in [0.5001,\,0.5480]$. 
    (c) Predictions for fixed $c=0.3099$ and $s=0.5121$, with $\gamma$ varied over the indicated values. 
    In all panels, the endpoints of each theoretical trajectory correspond to the number of e-folds 
    $N = 50$ (small markers) and $N = 60$ (large markers).
    }
    \label{fig:combined_ns_alphas_numerical}
\end{figure}
For convenience, we introduce the auxiliary variable (setting
$M_{Pl}=1$)
\be
Y(N)
=
2\xi(2s-1)\,\phi(N)^{c}\,T(N)^{\,s-1}, 
\ee such that the effective gravitational coupling can be written as
\be
\mathcal{G}_{eff}(N)
=
\frac{\mathcal{G}}{1+Y(N)}.
\ee  In the limit $Y \gg 1$, the theory enters the dominant-coupling regime discussed in the previous section.

\begin{table*}[t]
\centering
\caption{Numerical predictions for inflationary observables and derived model parameters. The table is divided into three blocks corresponding to variations in $c$ (top), $s$ (middle), and $\gamma$ (bottom). Observables are reported for $N=50$ and $N=60$ e-folds, while derived parameters are evaluated at $N=60$(horizon crossing).}
\label{tab:model_results_numerical}
\renewcommand{\arraystretch}{1.3} 
\resizebox{\textwidth}{!}{%
\begin{tabular}{!{\vrule width 1.5pt}c|c|c!{\vrule width 1.5pt}c|c|c!{\vrule width 1.5pt}c|c|c!{\vrule width 1.5pt}c|c|c|c|c|c!{\vrule width 1.5pt}c}
\noalign{\hrule height 1.5pt}

\multicolumn{3}{!{\vrule width 1.5pt}c!{\vrule width 1.5pt}}{\textbf{Input}} & 
\multicolumn{3}{c!{\vrule width 1.5pt}}{\textbf{Observables ($N=50$)}} & 
\multicolumn{3}{c!{\vrule width 1.5pt}}{\textbf{Observables ($N=60$)}} & 
\multicolumn{6}{c!{\vrule width 1.5pt}}{\textbf{Derived Parameters ($N=60$)}} \\
\noalign{\hrule height 1.0pt} 
$c$ & $s$ & $\gamma$ & $n_s$ & $r$ & $\alpha_s$ & $n_s$ & $r$ & $\alpha_s$ & $\xi$ & $\lambda$ & $\epsilon$ & $\eta$ & $H$ & $E_{\text{inf}}$  \\

\noalign{\hrule height 1.5pt}
0.2150 & 0.5121 & $10^{-10}$ & 0.9767 & 0.0237 & -4.88e-04 & 0.9767 & 0.0203 & -3.37e-04 & 0.0041 & 1.27e-15 & 0.0103 & 0.1516 & 9.29e-05 & 0.0453  \\
0.2737 & 0.5121 & $10^{-10}$ & 0.9744 & 0.0274 & -5.29e-04 & 0.9744 & 0.0234 & -3.66e-04 & 0.0038 & 1.08e-15 & 0.0104 & 0.1350 & 1.05e-04 & 0.0480  \\
0.3325 & 0.5121 & $10^{-10}$ & 0.9726 & 0.0312 & -5.62e-04 & 0.9726 & 0.0265 & -3.90e-04 & 0.0035 & 9.25e-16 & 0.0105 & 0.1228 & 1.18e-04 & 0.0510  \\
0.3912 & 0.5121 & $10^{-10}$ & 0.9710 & 0.0351 & -5.89e-04 & 0.9710 & 0.0297 & -4.09e-04 & 0.0033 & 8.09e-16 & 0.0106 & 0.1135 & 1.34e-04 & 0.0543  \\
0.4499 & 0.5121 & $10^{-10}$ & 0.9696 & 0.0391 & -6.14e-04 & 0.9696 & 0.0330 & -4.26e-04 & 0.0031 & 7.18e-16 & 0.0107 & 0.1064 & 1.52e-04 & 0.0579  \\

\noalign{\hrule height 1.5pt}
0.3099 & 0.5001 & $10^{-10}$ & 0.9731 & 0.0294 & -5.52e-04 & 0.9731 & 0.0250 & -3.83e-04 & 0.0035 & 9.53e-16 & 0.0104 & 0.1226 & 1.12e-04 & 0.0496  \\
0.3099 & 0.5121 & $10^{-10}$ & 0.9740 & 0.0329 & -5.32e-04 & 0.9740 & 0.0279 & -3.69e-04 & 0.0043 & 1.17e-15 & 0.0108 & 0.1553 & 1.20e-04 & 0.0514  \\
0.3099 & 0.5241 & $10^{-10}$ & 0.9746 & 0.0371 & -5.16e-04 & 0.9746 & 0.0313 & -3.59e-04 & 0.0052 & 1.39e-15 & 0.0111 & 0.1785 & 1.28e-04 & 0.0531  \\
0.3099 & 0.5360 & $10^{-10}$ & 0.9749 & 0.0415 & -5.06e-04 & 0.9749 & 0.0350 & -3.52e-04 & 0.0061 & 1.61e-15 & 0.0113 & 0.1953 & 1.35e-04 & 0.0546  \\
0.3099 & 0.5480 & $10^{-10}$ & 0.9750 & 0.0461 & -5.02e-04 & 0.9750 & 0.0389 & -3.49e-04 & 0.0072 & 1.83e-15 & 0.0115 & 0.2076 & 1.42e-04 & 0.0560  \\

\noalign{\hrule height 1.5pt}
0.3099 & 0.5121 & $10^{-12}$ & 0.9710 & 0.0254 & -5.83e-04 & 0.9710 & 0.0213 & -4.06e-04 & 0.0263 & 5.77e-16 & 0.0110 & 0.1216 & 2.24e-04 & 0.0703  \\
0.3099 & 0.5121 & $10^{-11}$ & 0.9719 & 0.0269 & -5.71e-04 & 0.9719 & 0.0227 & -3.97e-04 & 0.0094 & 7.03e-16 & 0.0108 & 0.1227 & 1.56e-04 & 0.0586  \\
0.3099 & 0.5121 & $10^{-10}$ & 0.9732 & 0.0297 & -5.50e-04 & 0.9732 & 0.0253 & -3.81e-04 & 0.0036 & 9.78e-16 & 0.0104 & 0.1271 & 1.13e-04 & 0.0498  \\
0.3099 & 0.5121 & $10^{-9}$  & 0.9749 & 0.0350 & -5.21e-04 & 0.9749 & 0.0300 & -3.60e-04 & 0.0015 & 1.57e-15 & 0.0103 & 0.1383 & 8.57e-05 & 0.0435  \\
0.3099 & 0.5121 & $3\times10^{-9}$ & 0.9757 & 0.0388 & -5.05e-04 & 0.9757 & 0.0335 & -3.49e-04 & 0.0010 & 2.05e-15 & 0.0103 & 0.1470 & 7.67e-05 & 0.0411  \\

\noalign{\hrule height 1.5pt}
\end{tabular}%
}
\end{table*}

This definition allows us to rewrite the background equations in a compact
form. Assuming the Higgs potential \eqref{Higgs} in the large-field regime
$\phi^{2}\gg\nu^{2}$—or, more generally, a monomial potential of the form $V(\phi)=\lambda \phi^{d}/d$ with $d=4$—together with the coupling function \eqref{Coupling_F},
the first Friedmann equation in the slow-roll approximation can be expressed
as
\be
\left(2(2s-1)\right)^{\frac{1}{1-s}}
(Y+1)\,Y^{\frac{1}{s-1}}\,
\phi^{\frac{c}{1-s}}
-
2\gamma\,\phi^{d}
=
0,
\label{FrEQFull}
\ee
where we have defined the parameter
\be
\gamma \equiv \frac{\lambda}{d\: \xi^{\frac{1}{1-s}}}.
\ee
Equation~\eqref{FrEQFull} can be solved algebraically for the scalar field as a
function of $Y(N)$, yielding
\be
\phi(N)
=
\left[
\frac{2\gamma\,Y(N)^{\frac{1}{1-s}}}
{\left(2(2s-1)\right)^{\frac{1}{1-s}}\left(Y(N)+1\right)}
\right]^{\frac{1}{\frac{c}{1-s}-d}}.
\label{phi_N_full}
\ee

On the other hand, the equation of motion for the scalar field, written in
terms of the number of e-folds and evaluated under the slow-roll
approximation, takes the form
\be
\frac{d\phi}{dN}
=
\frac{
\gamma d \left(2(2s-1)\right)^{\frac{s}{s-1}}
Y(N)^{\frac{1}{1-s}}
\phi(N)^{\frac{c}{s-1}+d}
+
c\,Y(N)
}
{(2s-1)\,\phi(N)}.
\label{kleinfull}
\ee

Taking the derivative of 
\eqref{FrEQFull} and using \eqref{kleinfull} and \eqref{phi_N_full}, we obtain 
{\small
\be
\frac{dY}{dN}=\frac{\mathcal{A}  Y^{1-\frac{2}{c+d (s-1)}} (Y+1)^{\frac{c+(d-2) (s-1)}{c+d (s-1)}}\left(c 2^{\frac{s}{s-1}} Y+d (4 s-2)^{\frac{s}{s-1}} (2 s-1)^{\frac{1}{1-s}} (Y+1)\right)}{s Y+1},
\label{full_Y_N}
\ee} where 
\be
\mathcal{A}=2^{-\frac{s (c+(d-2) (s-1))}{(s-1) (c+d (s-1))}} \gamma^{\frac{2 (s-1)}{c+d (s-1)}} (c+d (s-1)) (2 s-1)^{\frac{2}{c+d (s-1)}-1}.
\ee is a constant.  Equations~\eqref{phi_N_full} and \eqref{full_Y_N} form a closed system that can
be integrated numerically to determine the full inflationary evolution without relying on the dominant-coupling (high-energy) approximation.



The scalar power spectrum is given by
\be
\mathcal{P}_{s}(N_{*})=\frac{2^{-\frac{3}{s-1}-5} (2 s-1)^{-\frac{3}{s-1}} \xi ^{\frac{1}{1-s}} \phi_{*}^{2-\frac{c}{s-1}} Y_{*}^{\frac{3}{s-1}}}{3 \pi ^2 \left(d \gamma \phi^{\frac{c}{s-1}+d}+c (4 s-2)^{\frac{s}{1-s}} Y_{*}^{\frac{s}{s-1}}\right)^2}, 
\ee where $\phi(N_{*})\equiv\phi_{*}$, $Y(N_{*})\equiv Y_{*}$.

Thus, solving for $\xi$ we get
\be
\xi(N_{*})=\frac{2^{2-5 s} 3^{1-s} \pi ^{2-2 s} Y_{*}^3 \mathcal{P}_{s}^{1-s} \phi_{*}^{-c+2 s-2} \left(d \gamma \phi_{*}^{\frac{c}{s-1}+d}+c (4 s-2)^{\frac{s}{1-s}} Y_{*}^{\frac{s}{s-1}}\right)^{2-2 s}}{(2 s-1)^3}, 
\ee  and

\be
\lambda(N_{*})=\gamma d\:\xi(N_{*})^{\frac{1}{1-s}}.
\ee


\begin{figure}[!b]
    \centering
    \begin{subfigure}[b]{0.40\textwidth}
        \centering
        \includegraphics[width=\textwidth]{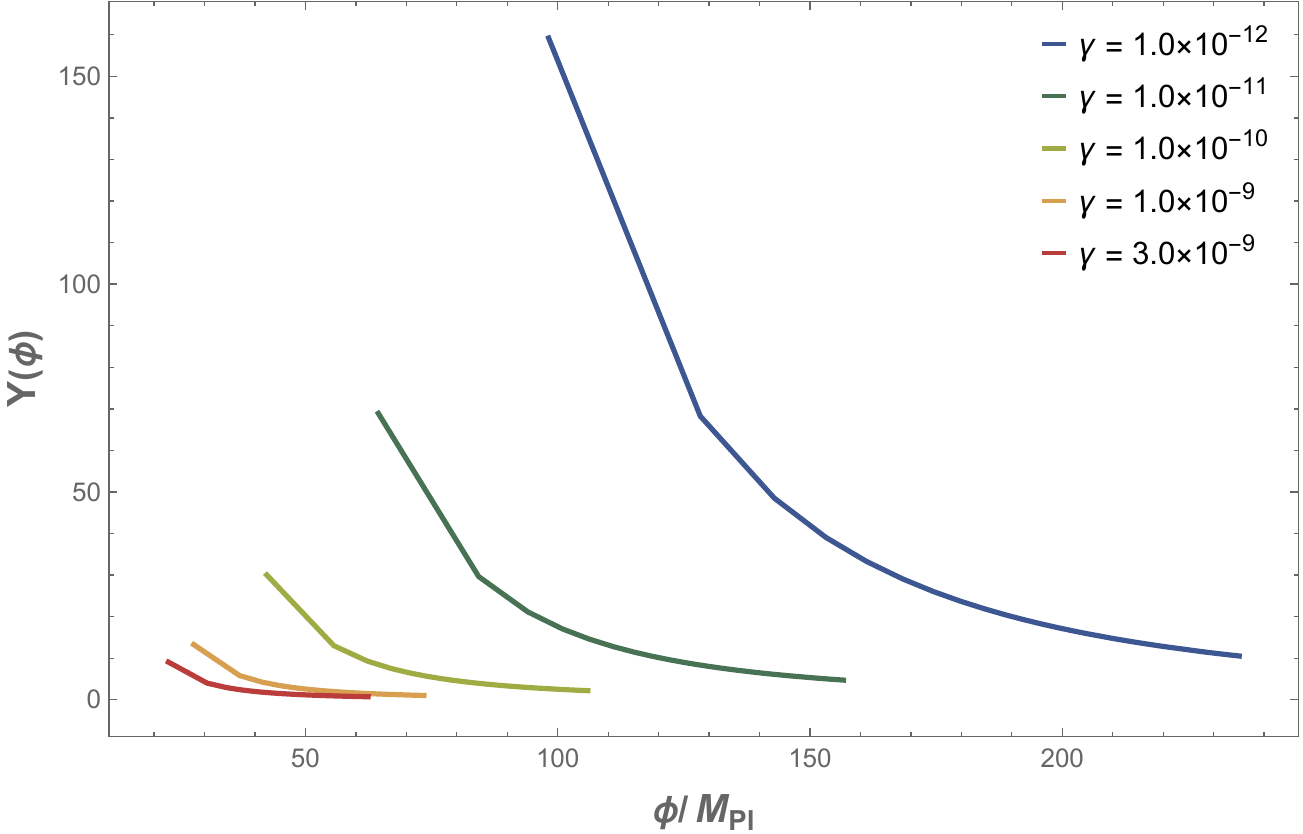}
        \caption{}
        \label{fig:Y_vs_phi}
    \end{subfigure}
    \hspace{1cm} 
    \begin{subfigure}[b]{0.46\textwidth}
        \centering
        \includegraphics[width=\textwidth]{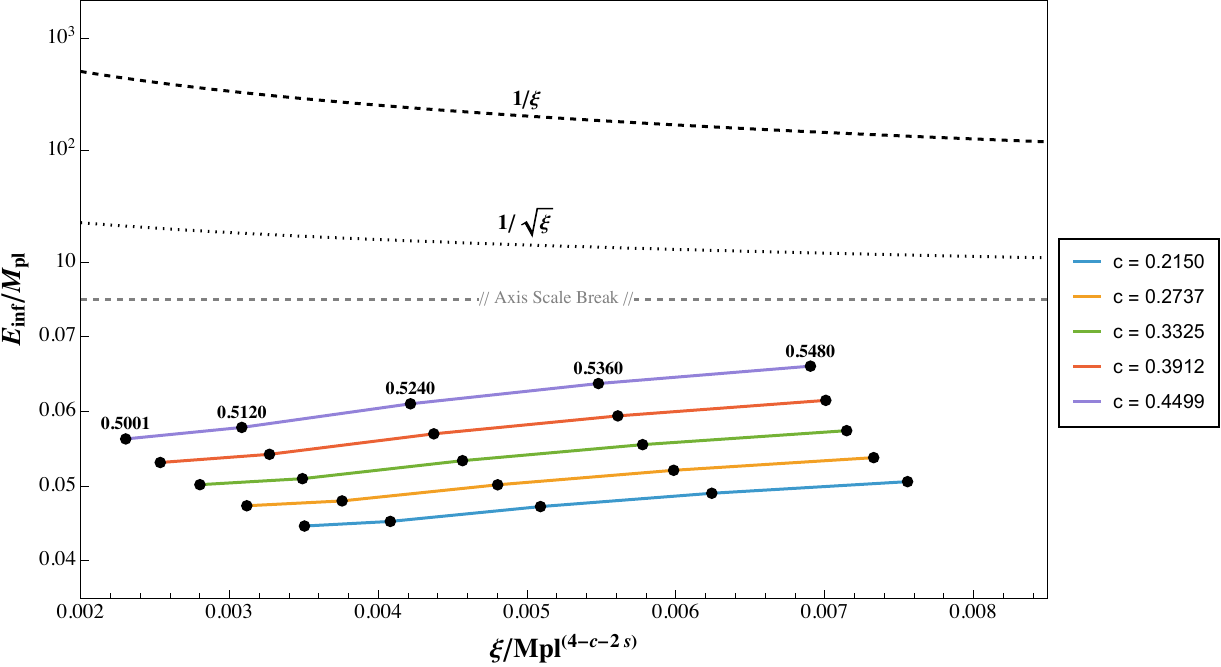}
        \caption{}
        \label{fig:Einf_vs_xi}
    \end{subfigure}
    \caption{Illustration of the coupling regime and the relation between the inflationary energy scale and the effective non-minimal coupling in Higgs-like inflation with $f(T,\phi)$ gravity.\textbf{(a)} Evolution of the auxiliary variable $Y(\phi)$ as a function of the scalar field $\phi$ for several values of the parameter $\gamma$, with $s=0.511$ and $c=0.3119$ fixed. Smaller values of $\gamma$ lead to larger values of $Y(\phi)$, driving the system toward the dominant-coupling (high-energy) regime in which $Y\gg 1$.\textbf{(b)} Inflationary energy scale $E_{\rm inf}$ as a function of the coupling parameter $\xi$ for representative values of $c$ and $s$. The dashed and dotted curves indicate the reference scales $1/\xi$ and $1/\sqrt{\xi}$, respectively, commonly used as estimates of the perturbative cutoff in non-minimally coupled inflationary models. We have used two different order scales on the y-axis to accommodate both the plots of the $(\xi-E_{\rm inf})$ plane and reference scales $1/\xi$ and $1/\sqrt{\xi}$, separated by the dashed line. The black dots marked on 
each curve corresponds to specific values of the parameter $s$; the $s$ values 
are explicitly labeled on the uppermost curve ($c = 0.4499$) as 
$s = 0.5001,\, 0.5120,\, 0.5240,\, 0.5360,\, 0.5480$, and each black dot on 
the lower curves ($c = 0.3912,\, 0.3325,\, 0.2737,\, 0.2150$) corresponds to 
the same value of $s$ as the corresponding aligned dot on the uppermost curve. The inflationary trajectories remain below these reference scales throughout the considered parameter region. The two panels together illustrate the connection between the coupling regime characterized by $Y(\phi)$ and the corresponding inflationary energy scale, showing that the parameter space consistent with observations remains within the expected effective-field-theory validity range.
}
\end{figure}

\begin{figure}[!b]
    \centering
    \includegraphics[width=0.85\textwidth, height=0.60\textheight]{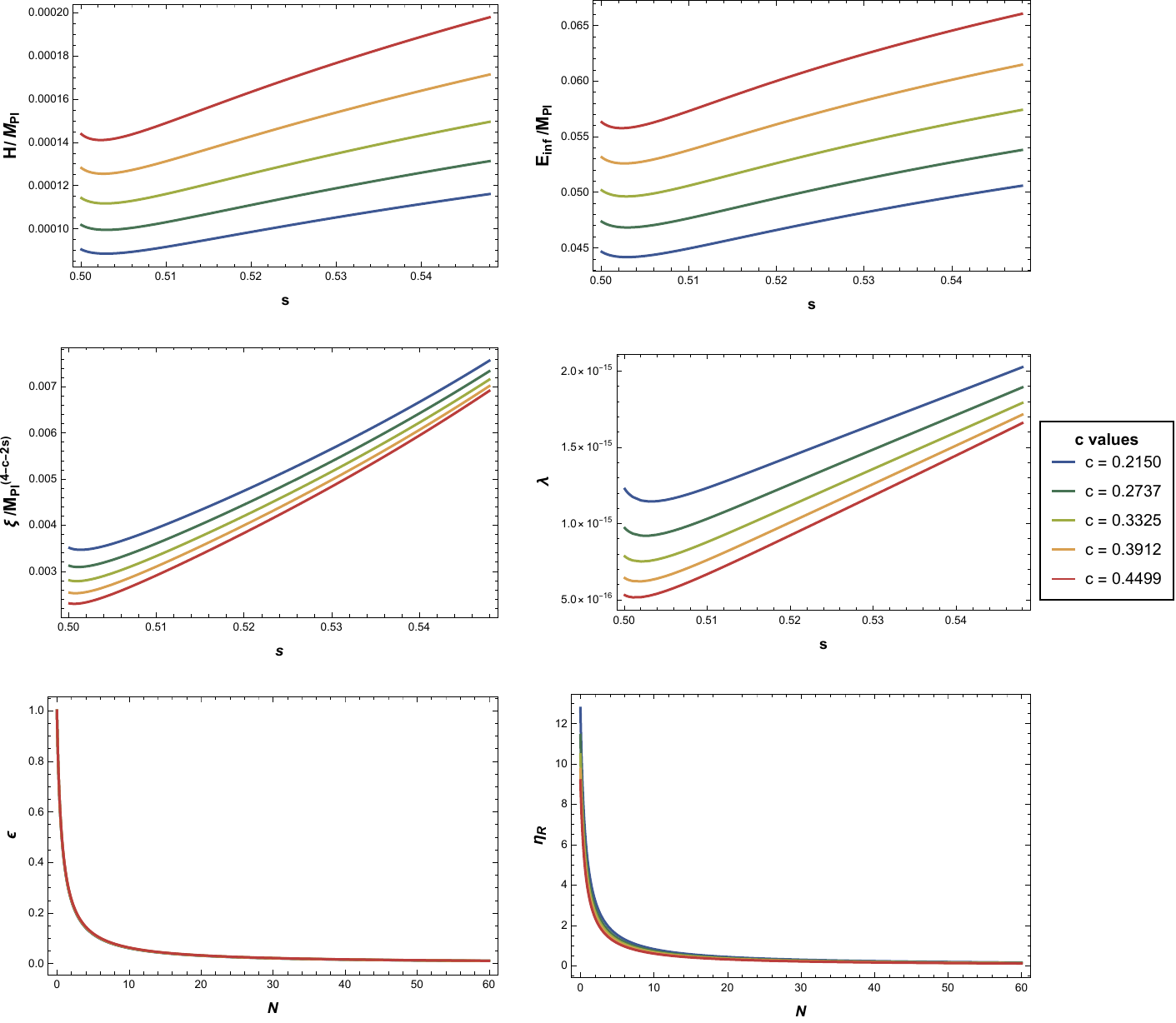}
    \caption{Background inflationary quantities in the Higgs-like inflation 
model with $f(T,\phi)$ gravity, obtained from the numerical integration of the slow-roll equations beyond the dominant-coupling approximation, with $\gamma = 10^{-10}$ held fixed throughout to remain away from the high-energy regime (cf. Sec.~\ref{Strong_Coupling}). \textit{Upper panels}: the Hubble parameter $H(N_\ast)/M_{\rm Pl}$, the inflationary energy scale $E_{\rm inf}/M_{\rm Pl}$, the non-minimal coupling $\xi(N_\ast)/M_{\rm Pl}^{4-c-2s}$, and the effective 
self-coupling $\lambda(N_\ast)$, all evaluated at horizon crossing $N_\ast = 60$ and displayed as functions of $s$ for $s \in [0.5001,\,0.5480]$. \textit{Lower panels}: the slow-roll parameters $\epsilon(N)$ and $\eta_R(N)$ as functions of the number of e-folds $N$, computed at the fixed representative value $s = 0.5121$ over the full inflationary interval $N \in [0,\,60]$. Each curve 
corresponds to a distinct value of $c \in [0.2150,\,0.4499]$, illustrating how the torsion-scalar coupling modifies the inflationary dynamics across the explored parameter range.}
    \label{fig:all_parameter_plots_numerical_sol}
\end{figure}

The spectral index takes the form
\bea
&& n_{s}(N_{*})=1+\frac{2 (c-1) c Y_{*}}{(2 s-1) \phi_{*} ^2}+(d-1) d \gamma 2^{\frac{1}{s-1}+2} (2 s-1)^{\frac{1}{s-1}} Y_{*}^{\frac{1}{1-s}} \phi_{*} ^{\frac{c}{s-1}+d-2}-\nonumber\\
&& \frac{c d \gamma 2^{\frac{1}{1-s}+2} (2 s-1)^{-\frac{s}{s-1}} (4 s-2)^{\frac{2}{s-1}} Y_{*}^{\frac{s-2}{s-1}} (s (3 s-1) Y_{*}+s (7-2 s)-3) \phi_{*} ^{\frac{c}{s-1}+d-2}}{(s Y_{*}+1) (s Y_{*}+2 s-1)}+\nonumber\\
&& 
\frac{c^2 Y_{*} \left[Y_{*} \left(3+s \left(-4 s^2+2 s-5\right)-s \left(2 s^2+s-1\right) Y_{*}\right)+2 s (1-2 s)\right]}{(s-1) (2 s-1) \phi_{*} ^2 (s Y_{*}+1) (s Y+2 s-1)}-\nonumber\\
&& \frac{2 d^2 \gamma^2 (2 s-1)^{\frac{1}{s-1}} Y_{*}^{-\frac{2}{s-1}} \left[s \left(2^{\frac{1}{s-1}+3} s-2^{\frac{s}{s-1}}\right) (4 s-2)^{\frac{1}{s-1}} Y_{*}+3\ 2^{\frac{s+1}{s-1}} (2 s-1)^{\frac{s}{s-1}}\right] \phi_{*} ^{2 \left(\frac{c}{s-1}+d-1\right)}}{(s Y_{*}+1) (s Y_{*}+2 s-1)},\nonumber\\
\eea
and its running is given by
\[
\begin{aligned}
\alpha_s(N_*) =\;
\frac{1}
{(s-1)(2s-1)\,(1+sY_*)^{3}\,(2s-1+sY_*)^{2}\,\phi_*^{3}}  \sum_{i=1}^{11}\mathcal{F}_i(\phi_*,Y_*) .
\end{aligned}
\]
where $\mathcal{F}_i(\phi_*,Y_*)$ functions are 
defined in appendix \ref{running_beyond}.

On the other hand, the tensor-to-scalar ratio reads 
\be
r(N_{*})=\frac{2^{\frac{2}{s-1}+5} (2 s-1)^{\frac{s+1}{s-1}} Y_{*}^{-\frac{2}{s-1}} \left(d \gamma \phi_{*} ^{\frac{c}{s-1}+d}+c (4 s-2)^{\frac{s}{1-s}} Y_{*}^{\frac{s}{s-1}}\right)^2}{\phi_{*} ^2 (s (Y_{*}+2)-1)}.
\ee

In Fig.~\ref{fig:combined_ns_r_numerical}, we present the parametric trajectories of the scalar spectral index $n_{s}$ versus the tensor-to-scalar ratio $r$. These theoretical predictions are obtained by numerically integrating the slow-roll equations without assuming the dominant-coupling (high-energy) approximation. We explore the parameter space by varying $c \in [0.2150, 0.4499]$ and $s \in [0.5001, 0.5480]$, alongside the coupling parameter $\gamma$, for the number of e-folds $50 \leq N_{*} \leq 60$. The results demonstrate that, even outside the high-energy limit, the model predictions remain robust and lie well within the $1\sigma$ and $2\sigma$ confidence regions of the combined dataset from Planck 2018, ACT DR6 (ACT-2025), and BICEP/Keck 2021.

  Fig.~\ref{fig:combined_ns_alphas_numerical} presents the marginalized constraints in the $n_{s}-\alpha_{s}$ plane. The theoretical predictions for the running of the scalar spectral index are consistent with the $68\%$ and $95\%$ confidence intervals derived from the joint Planck, ACT, and DESI analysis. Table~\ref{tab:model_results_numerical} summarizes these findings, listing the specific values for $n_{s}$, $r$, and $\alpha_{s}$, along with the derived model parameters at $N=60$. The table confirms that the numerical results align with current observational bounds. Furthermore, Fig.~\ref{fig:Y_vs_phi} illustrates the role of the coupling function $Y(\phi)$ (or $\mathcal{G}_{eff}(\phi)$) across different energy scales. 
In particular, smaller values of the parameter $\gamma$ drive the system toward the high-energy approximation, characterized by large values of both $Y(\phi)$ and the inflaton field $\phi$, reaching trans-Planckian field values. 
In this regime, the coupling term dominates the dynamics, placing the model firmly in the dominant-coupling regime and providing a smooth connection with the results obtained in the previous section. Conversely, increasing $\gamma$ shifts the dynamics toward lower values of $Y(\phi)$ and $\phi$. 
However, this regime is strongly constrained by observational data, since larger values of $\gamma$ lead to an enhanced tensor-to-scalar ratio $r$, pushing the predictions outside the $2\sigma$ confidence region. We display in Fig.~\ref{fig:Einf_vs_xi} the inflationary energy scale
$E_{\rm inf}/M_{\rm Pl}$ as a function of the nonminimal scalar-torsion
coupling $\xi$. The reference scales
$\Lambda_1\sim M_{\rm Pl}/\xi$ and
$\Lambda_2\sim M_{\rm Pl}/\sqrt{\xi}$ are also shown. These scales are
commonly used as cutoff estimates in curvature-based Higgs inflation with
nonminimal coupling $\xi\phi^2R$. As discussed above, they are not identified
with the perturbative cutoff of the present scalar-torsion theory, where the
nonminimal interaction is given by $F(\phi)G(T)$, with
$F(\phi)=\xi\phi^c$ and $G(T)\sim T^s$. The comparison is therefore used
only to indicate the position of the inflationary energy scale relative to
the standard Higgs-inflation benchmark scales. For the parameter region
considered here, the predicted inflationary energy scale remains below both $\Lambda_1$ and $\Lambda_2$, namely
$E_{\rm inf}<\Lambda_1,\Lambda_2$ .
Thus, the model is not driven into an energy regime that would be in
immediate tension with these reference cutoff estimates. Nevertheless, since
these scales are not the actual cutoff scales of the scalar-torsion theory,
this comparison should be regarded only as a consistency check at the level
of benchmark scales, not as a proof of perturbative unitarity.

Finally,  Fig.~\ref{fig:all_parameter_plots_numerical_sol} illustrates the evolution of the key background quantities, i.e. the Hubble parameter $H(N_{*})$, the inflationary energy scale $E_{\text{inf}}$, the non-minimal coupling $\xi(N_{*})$, and the effective self-coupling $\lambda(N_{*})$, evaluated at $N_{*}=60$ and outside the high energy regime. It also displays the slow-roll parameters $\epsilon$ and $\eta_{\mathcal{R}}$ throughout the inflationary epoch. The smooth variation of the background fields and the fact that the slow-roll parameters remain much smaller than unity ($\epsilon, |\eta_{\mathcal{R}}| \ll 1$) confirm the dynamical stability of the inflationary solution across the explored parameter range.


\section{Conclusions}
\label{Conclusions}

In this work we have investigated Higgs-like inflation within the framework
of scalar-torsion gravity, focusing on the general class of $f(T,\phi)$
theories. The motivation for this study is twofold. On the one hand, Higgs-like
inflation provides an attractive bridge between high-energy physics and the
early Universe, but its viability is increasingly scrutinized by the steadily
improving precision of cosmological observations. On the other hand,
torsion-based extensions of gravity constitute a conceptually distinct
alternative to curvature-based theories, giving rise to modified gravitational interactions and cosmological dynamics. In view of the latest
CMB and large-scale structure constraints from Planck, ACT, DESI, and
BICEP/Keck, it is therefore timely to reassess Higgs-like inflationary
scenarios within a scalar-torsion gravitational setting.

We first developed the theoretical framework of $f(T,\phi)$ gravity at the
background and perturbative levels, emphasizing the role of scalar-torsion
interactions and the consequences of local Lorentz-symmetry breaking for the
inflationary dynamics. An important aspect of our analysis is the use of the scalar and tensor power spectra derived within the $f(T,\phi)$ framework in
Refs.~\cite{Gonzalez-Espinoza:2020azh,Gonzalez-Espinoza:2021qnv}. This
distinguishes the present treatment from previous scalar-torsion inflationary
studies~\cite{Chakrabortty2021Inflation,Chakrabortty2022Correction,
Fomin2025ScalarTorsion}, where the inflationary observables were evaluated using the standard GR-based spectra and consistency relations as an approximation. In
$f(T,\phi)$ gravity, the breaking of local Lorentz symmetry introduces
additional degrees of freedom, associated with the Goldstone modes of the
broken symmetry, which modify the perturbation sector. Therefore, the use of
the scalar and tensor spectra derived specifically for scalar-torsion gravity
is essential for obtaining the inflationary predictions consistently. Within the slow-roll approximation, we derived the background equations, the
second-order actions for scalar and tensor perturbations, and the
corresponding inflationary observables. We then focused on the
dominant-coupling, or high-energy, regime, in which the scalar-torsion
interaction controls the inflationary dynamics. In this limit, we obtained
closed-form analytical expressions for the scalar spectral index $n_s$, the
tensor-to-scalar ratio $r$, and the running of the scalar spectral index
$\alpha_s$ as functions of the number of e-folds. These expressions show how
torsion modifies the inflationary observables and leads to consistency
relations that differ from those of standard single-field inflation.

Specializing to Higgs-like inflation, we considered a quartic symmetry-breaking potential together with a power-law scalar-torsion coupling. In the dominant-coupling regime, the model predicts values of $n_s$, $r$, and $\alpha_s$ compatible with current observational bounds, including the joint constraints from Planck, ACT, DESI, and BICEP/Keck. In particular, the scalar-torsion corrections shift the predictions toward the slightly larger values of $n_s$ preferred by recent ACT and DESI data, while remaining consistent with the upper bound on the tensor-to-scalar ratio. We also obtained a direct parametric relation $r(n_s)$, independent of the number of e-folds, and explicit expressions for the effective self-coupling and nonminimal coupling parameters. These results clarify the region of parameter space in which Higgs-like inflation in $f(T,\phi)$ gravity remains observationally viable.

To assess the robustness of the analytical results, we performed a numerical
analysis beyond the dominant-coupling approximation, while remaining within
the slow-roll regime. The slow-roll system was integrated by introducing the
auxiliary variable $Y(N)$, which parametrizes the effective gravitational
coupling and allows the evolution to be tracked in the parameter space $(c,s,\gamma)$. In this parametrization, $\gamma$ controls the transition
between the dominant-coupling regime and the regime in which the
scalar-torsion interaction becomes subdominant. This procedure provides a systematic way to determine how the analytical high-energy predictions are modified when the dominant-coupling approximation is relaxed.

The numerical results show that the viable regions identified analytically persist beyond the dominant-coupling limit. In particular, Higgs-like inflation in $f(T,\phi)$ gravity remains compatible with current observational constraints on $n_s$, $r$, and $\alpha_s$, while exhibiting characteristic modifications in the tensor sector and in the scale dependence of the scalar spectrum. The running of the scalar spectral index was computed both analytically in the dominant-coupling regime and numerically beyond this approximation. In the numerical treatment, the corresponding auxiliary functions $F_i(\phi_*,Y_*)$ are given in Appendix~B. This is relevant in view of recent CMB analyses, in particular ACT-based constraints, which have increased the phenomenological interest in the scale dependence of the scalar spectrum. Thus, the computation of $\alpha_s$ for the Higgs-like scalar-torsion model considered here provides an additional element in the comparison with current data, complementing the predictions for $n_s$ and $r$.

Since the inflaton is treated as a generic Higgs-like scalar and the nonminimal interaction is scalar-torsion rather than curvature-based, the standard cutoff estimates $\Lambda_1\sim M_{\rm Pl}/\xi$ and $\Lambda_2\sim M_{\rm Pl}/\sqrt{\xi}$, which arise in Standard Model Higgs inflation with a nonminimal coupling $\xi\phi^2R$
\cite{Burgess:2009ea,Barbon:2009ya}, do not determine the actual cutoff
scale of the present theory. Nevertheless, they provide standard benchmark scales for comparison with the conventional Higgs-inflation scenario. Accordingly, Fig.~\ref{fig:Einf_vs_xi} compares the inflationary
energy scale predicted by the scalar-torsion model with these reference
scales. In the parameter region explored, the predicted value of
$E_{\rm inf}$ remains below both $\Lambda_1$ and $\Lambda_2$. This shows that, within the explored parameter region, the inflationary scale lies below the standard Higgs-inflation benchmark scales. However, this comparison should be regarded only as a reference consistency check, not as a determination of the perturbative cutoff or as a proof of perturbative unitarity. The actual cutoff scale of the scalar-torsion effective theory must be obtained from a
dedicated perturbative analysis, by expanding the action around the relevant inflationary background and identifying the higher-dimensional operators that control scattering amplitudes \cite{Hu:2023juh,Hu:2023xcf}.

In conclusion, our results show that $f(T,\phi)$ gravity provides a viable framework in which Higgs-like inflation can
be reconciled with precision cosmology. The characteristic inflationary signatures induced by scalar-torsion couplings could be tested by upcoming CMB polarization experiments and next-generation large-scale-structure
surveys. Extensions of the present analysis to include reheating, primordial
black-hole formation, or non-Gaussianities represent promising directions
and will be studied in future works.


\begin{acknowledgments}
E.N.S. gratefully acknowledges  the 
contribution of 
the LISA Cosmology Working Group (CosWG), as well as support from the COST 
Actions CA21136 -  Addressing observational tensions in cosmology with 
systematics and fundamental physics (CosmoVerse)  - CA23130, Bridging 
high and low energies in search of quantum gravity (BridgeQG)  and CA21106 -  
 COSMIC WISPers in the Dark Universe: Theory, astrophysics and 
experiments (CosmicWISPers). M.G.-E. acknowledges the financial support of FONDECYT de Postdoctorado, No. 3230801. N.K. gratefully
acknowledges the scholarship support provided by the Consortium Doctoral Program in Physics jointly oﬀered by Universidad de Valparaíso,
Universidad de Tarapacá, and Universidad de La Serena. R.R. and B.E.
acknowledge financial support through scholarships from the Master's
Program in Physics at the University of Tarapacá.
\end{acknowledgments}

\begin{appendix}
\section{ The running of the scalar spectral index  \texorpdfstring{$\alpha_s$}{alpha s}}\label{func_alpha}

In the scenario at hand, the running of the scalar spectral index is given by 
\begin{equation}
    \alpha_s \equiv \frac{dn_s}{d\ln k} = \frac{\sqrt{6}}{T^{7/2}} 
    \left[\Theta_1 + 4(\Theta_2 + \Theta_3)+ \Theta_4 \right],
\end{equation}
where $\Theta_1$, $\Theta_2$, $\Theta_3$ and $\Theta_4$ are 
given by  
\be
\begin{aligned}
\Theta_{1}\equiv{}\;&
\frac{4\,(G F_{,\phi}+V_{,\phi})\big(F_{,\phi}(G-2T G_{,T})+V_{,\phi}\big)}
{M_{Pl}^{2}+2F(G_{,T}+2T G_{,TT})}
\\[4pt]
&\times
\Bigg[
-2\dot T
+\frac{T\big(F_{,\phi}G_{,T}\dot T+\dot\phi(GF_{,\phi\phi}+V_{,\phi\phi})\big)}
{G F_{,\phi}+V_{,\phi}}
\\
&\qquad
+\frac{T\big(-F_{,\phi}\dot T(G_{,T}+2T G_{,TT})
+\dot\phi\big((G-2T G_{,T})F_{,\phi\phi}+V_{,\phi\phi}\big)\big)}
{F_{,\phi}(G-2T G_{,T})+V_{,\phi}}
\\
&\qquad
-\frac{2T\big(F_{,\phi}\dot\phi(G_{,T}+2T G_{,TT})
+F\dot T(3G_{,TT}+2T G_{,TTT})\big)}
{M_{Pl}^{2}+2F(G_{,T}+2T G_{,TT})}
\Bigg],
\end{aligned}
\ee
\be
\begin{aligned}
\Theta_{2}\equiv{}\;&
-\frac{2T^{2}F_{,\phi}^{2}G_{,T}\dot T}{F}
+\frac{4\dot T\,(G F_{,\phi}+V_{,\phi})^{2}}{M_{Pl}^{2}+2F G_{,T}}
+\frac{T\,F_{,\phi}^{2}G_{,T}^{2}\dot T}{F\,G_{,TT}}
+\frac{T^{2}F_{,\phi}^{3}G_{,T}^{2}\dot\phi}{F^{2}G_{,TT}}
-\frac{2T^{2}F_{,\phi}G_{,T}^{2}\dot\phi\,F_{,\phi\phi}}{F\,G_{,TT}}
\\[4pt]
&\quad
+\frac{4T\,(G F_{,\phi}+V_{,\phi})^{2}
\big(F_{,\phi}G_{,T}\dot\phi+F\dot T\,G_{,TT}\big)}
{(M_{Pl}^{2}+2F G_{,T})^{2}}
\\[4pt]
&\quad
-\frac{4\dot T
\big(F_{,\phi}(G-2T G_{,T})+V_{,\phi}\big)
\big(F_{,\phi}(G-T G_{,T})+V_{,\phi}\big)}
{M_{Pl}^{2}+2F(G_{,T}+2T G_{,TT})}
\\[4pt]
&\quad
+\frac{2T\big(F_{,\phi}(G-2T G_{,T})+V_{,\phi}\big)
\Big((G-T G_{,T})\dot\phi\,F_{,\phi\phi}
-TF_{,\phi}\dot T\,G_{,TT}
+\dot\phi\,V_{,\phi\phi}\Big)}
{M_{Pl}^{2}+2F(G_{,T}+2T G_{,TT})},
\end{aligned}
\ee
{\footnotesize \be
\begin{aligned}
\Theta_{3}\equiv{}\;&
-\frac{4T\,(G F_{,\phi}+V_{,\phi})
\big(F_{,\phi}G_{,T}\dot T
+\dot\phi\,(G F_{,\phi\phi}+V_{,\phi\phi})\big)}
{M_{Pl}^{2}+2F G_{,T}}
\\[4pt]
&\quad
+\frac{2T\big(F_{,\phi}(G-T G_{,T})+V_{,\phi}\big)
\Big(-F_{,\phi}\dot T\,(G_{,T}+2T G_{,TT})
+\dot\phi\big((G-2T G_{,T})F_{,\phi\phi}+V_{,\phi\phi}\big)\Big)}
{M_{Pl}^{2}+2F(G_{,T}+2T G_{,TT})}
\\[4pt]
&\quad
+\frac{T^{2}F_{,\phi}^{2}G_{,T}^{2}\dot T\,G_{,TTT}}
{F\,G_{,TT}^{2}}
\\[4pt]
&\quad
-\frac{2T\big(F_{,\phi}(G-2T G_{,T})+V_{,\phi}\big)
\big(F_{,\phi}(G-T G_{,T})+V_{,\phi}\big)
\Big(2F_{,\phi}\dot\phi\,(G_{,T}+2T G_{,TT})
+2F\dot T\,(3G_{,TT}+2T G_{,TTT})\Big)}
{\big(M_{Pl}^{2}+2F(G_{,T}+2T G_{,TT})\big)^{2}},
\end{aligned}
\ee}
\be
\begin{aligned}
\!\!\!\!\!\!\!\!\!\!\!\!\!\!\!\!\!\!\!\!\!\!\!\!\!\!\!\!\!\!\!\!\!\!\!\!\!\!\!\!\!\!\!\!\!\!
\Theta_{4}\equiv{}\;&
\frac{1}{M_{Pl}^{2}+2F(G_{,T}+2T G_{,TT})}\;
\mathcal{A}\,\mathcal{B},
\end{aligned}
\ee
with
\be
\begin{aligned}
\mathcal{A}\equiv{}\;&
8G^{2}F_{,\phi}^{2}
+8\big(TF_{,\phi}G_{,T}-V_{,\phi}\big)\big(2TF_{,\phi}G_{,T}-V_{,\phi}\big)
\\
&\quad
-4G\Big(-4F_{,\phi}V_{,\phi}
+T\big(6F_{,\phi}^{2}G_{,T}+(M_{Pl}^{2}+2F G_{,T})F_{,\phi\phi}\big)
+4F T^{2}F_{,\phi\phi}G_{,TT}\Big)
\\
&\quad
-4T\big(M_{Pl}^{2}+2F(G_{,T}+2T G_{,TT})\big)V_{,\phi\phi},
\\[6pt]
\mathcal{B}\equiv{}\;&
-2\dot T
-\frac{2T\big(F_{,\phi}\dot\phi(G_{,T}+2T G_{,TT})
+F\dot T(3G_{,TT}+2T G_{,TTT})\big)}
{M_{Pl}^{2}+2F(G_{,T}+2T G_{,TT})},
\\
&\quad
+\frac{T}{\mathcal{D}}
\Big(
2F_{,\phi}G_{,T}\dot T\,V_{,\phi}
-4G^{2}F_{,\phi}\dot\phi\,F_{,\phi\phi}
+\big((M_{Pl}^{2}+2F G_{,T})\dot T-4V_{,\phi}\dot\phi\big)V_{,\phi\phi}
\\
&\qquad
+G\Big(
2F_{,\phi}^{2}\dot T(G_{,T}+3T G_{,TT})
+2F_{,\phi}\dot\phi\big(TF_{,\phi\phi}(7G_{,T}+2T G_{,TT})-2V_{,\phi\phi}\big)
\\
&\qquad\qquad
+\dot\phi\big(-4V_{,\phi}F_{,\phi\phi}
+T\big(M_{Pl}^{2}+2F(G_{,T}+2T G_{,TT})\big)F_{,\phi\phi\phi}\big)
\\
&\qquad\qquad
+\dot T\,F_{,\phi\phi}\big(M_{Pl}^{2}+2F(G_{,T}+5T G_{,TT}+2T^{2}G_{,TTT})\big)
\Big)
\\
&\qquad
+T\Big(
-2F_{,\phi}^{2}G_{,T}^{2}\dot T
+2F G_{,T}^{2}\dot T\,F_{,\phi\phi}
+10F\dot T\,G_{,TT}V_{,\phi\phi}
\\
&\qquad\qquad
+F_{,\phi}\big(6\dot T\,V_{,\phi}G_{,TT}
+8G_{,T}\dot\phi\,V_{,\phi\phi}\big)
+M_{Pl}^{2}\dot\phi\,V_{,\phi\phi\phi}
\\
&\qquad\qquad
+G_{,T}\big((M_{Pl}^{2}\dot T+6V_{,\phi}\dot\phi)F_{,\phi\phi}
+2F\dot\phi\,V_{,\phi\phi\phi}\big)
\Big)
\\
&\qquad
+4T^{2}\Big(
-2F_{,\phi}^{2}G_{,T}\dot T\,G_{,TT}
+F_{,\phi}\dot\phi\big(-2G_{,T}^{2}F_{,\phi\phi}+G_{,TT}V_{,\phi\phi}\big)
\\
&\qquad\qquad
+F\big(\dot T(G_{,T}F_{,\phi\phi}G_{,TT}+V_{,\phi\phi}G_{,TTT})
+\dot\phi\,G_{,TT}V_{,\phi\phi\phi}\big)
\Big)
\Big),
\\[6pt]
\mathcal{D}\equiv{}\;&
-2\,G^{2}F_{,\phi}^{2}
-2\big(TF_{,\phi}G_{,T}-V_{,\phi}\big)\big(2TF_{,\phi}G_{,T}-V_{,\phi}\big)
\\
&\quad
+G\Big(
-4F_{,\phi}V_{,\phi}
+T\big(6F_{,\phi}^{2}G_{,T}+(M_{Pl}^{2}+2F G_{,T})F_{,\phi\phi}\big)
+4F T^{2}F_{,\phi\phi}G_{,TT}
\Big)
\\
&\quad
+T\big(M_{Pl}^{2}+2F(G_{,T}+2T G_{,TT})\big)V_{,\phi\phi}.
\end{aligned}
\ee

\section{The running of the scalar spectral index \texorpdfstring{$\alpha_{s}(\phi_{*},Y_{*})$}{alpha s(phi*,Y*)} beyond the high-energy approximation} 
\label{running_beyond}
For the case beyond the high-energy approximation, the running of the scalar spectral index is given by
\be
\begin{aligned}
\alpha_s(\phi_*,Y_*) \equiv\;
\frac{1}
{(-1+s)(-1+2s)\,(1+sY_*)^{3}\,(-1+2s+sY_*)^{2}\,\phi_*^{3}}  \sum_{i=1}^{11}\mathcal{F}_i(\phi_*,Y_*) ,
\end{aligned}
\ee
where $\mathcal{F}_i(\phi_*,Y_*)$ functions are 
given by
\be
\begin{aligned}
\mathcal{F}_1(\phi_*,Y_*) \equiv\;&
2^{\frac{c(-2+s)+(-1+s)\bigl(d(-2+s)+s\bigr)}
{\bigl(c+d(-1+s)\bigr)(-1+s)}}\,
(c-1)c\,
\gamma^{\frac{-1+s}{c+d(-1+s)}}\,
(-1+s)\,
(-1+2s)^{-1+\frac{1}{c+d(-1+s)}}
\\[0.5em]
&\times
Y_*^{\,1-\frac{2}{-1+s}+\frac{1}{-c+d-ds}}\,
(1+Y_*)^{-\frac{c+(1+d)(-1+s)}{c+d(-1+s)}}\,
(1+sY_*)^{3}\,
(-1+2s+sY_*)^{2}
\\[0.5em]
&\times
\Bigl(Y_*^{\frac{1}{-1+s}}+Y_*^{\frac{s}{-1+s}}\Bigr)\,
\mathcal{K}_1(Y_*),
\end{aligned}
\ee
{\footnotesize \be
\begin{aligned}
\mathcal{F}_2(\phi_*,Y_*) \equiv\;&
-\,2^{-\frac{c+(d-s)(-1+s)}{\bigl(c+d(-1+s)\bigr)(-1+s)}}\,
\gamma^{\frac{-1+s}{c+d(-1+s)}}\,
(-1+2s)^{-1+\frac{1}{c+d(-1+s)}}\,
Y_*^{\,1-\frac{2}{-1+s}+\frac{1}{-c+d-ds}}
\\[0.5em]
&\times
(1+Y_*)^{-\frac{c+(1+d)(-1+s)}{c+d(-1+s)}}\,
(-1+2s+sY_*)\,
\bigl(c+c s Y_*\bigr)^{2}
\\[0.5em]
&\times
\Bigl(
2s(-1+2s)+(-3+5s-2s^{2}+4s^{3})Y_*+s(-1+s+2s^{2})Y_*^{2}
\Bigr)
\Bigl(Y_*^{\frac{1}{-1+s}}+Y_*^{\frac{s}{-1+s}}\Bigr)\,
\mathcal{K}_1(Y_*),
\end{aligned}
\ee}
\be
\begin{aligned}
\mathcal{F}_3(\phi_*,Y_*) \equiv\;&
2^{-\frac{s\bigl(c+(-2+d)(-1+s)\bigr)}{\bigl(c+d(-1+s)\bigr)(-1+s)}}\,
c^{2}\,
\gamma^{\frac{2(-1+s)}{c+d(-1+s)}}\,
\bigl(c+d(-1+s)\bigr)\,
(-1+2s)^{-1+\frac{2}{c+d(-1+s)}}
\\[0.5em]
&\times
Y_*^{\,2-\frac{2}{c+d(-1+s)}-\frac{2}{-1+s}}\,
(1+Y_*)^{\frac{2-2s}{c+d(-1+s)}}\,
(1+sY_*)\,(-1+2s+sY_*)
\\[0.5em]
&\times
\Bigl(-3+5s-2s^{2}+4s^{3}+2s(-1+s+2s^{2})Y_*\Bigr)
\Bigl(Y_*^{\frac{1}{-1+s}}+Y_*^{\frac{s}{-1+s}}\Bigr)\,
\mathcal{K}_1(Y_*)\;\phi_*,
\end{aligned}
\ee
\be
\begin{aligned}
\mathcal{F}_4(\phi_*,Y_*) \equiv\;&
2^{\,1-\frac{s\bigl(c+(-2+d)(-1+s)\bigr)}{\bigl(c+d(-1+s)\bigr)(-1+s)}}\,
(1-c)c\,
\gamma^{\frac{2(-1+s)}{c+d(-1+s)}}\,
\bigl(c+d(-1+s)\bigr)\,(1-s)\,
(-1+2s)^{-1+\frac{2}{c+d(-1+s)}}
\\[0.5em]
&\times
Y_*^{\,1-\frac{2}{c+d(-1+s)}-\frac{2}{-1+s}}\,
(1+Y_*)^{\frac{2-2s}{c+d(-1+s)}}\,
(1+sY_*)^{2}\,
(-1+2s+sY_*)^{2}
\\[0.5em]
&\times
\Bigl(Y_*^{\frac{1}{-1+s}}+Y_*^{\frac{s}{-1+s}}\Bigr)\,
\mathcal{K}_1(Y_*)\;\phi_*,
\end{aligned}
\ee
\be
\begin{aligned}
\mathcal{F}_5(\phi_*,Y_*) \equiv\;&
-\,2^{-\frac{s\bigl(c+(-2+d)(-1+s)\bigr)}{\bigl(c+d(-1+s)\bigr)(-1+s)}}\,
c^{2}\,
\gamma^{\frac{2(-1+s)}{c+d(-1+s)}}\,
\bigl(c+d(-1+s)\bigr)\,s\,
(-1+2s)^{-1+\frac{2}{c+d(-1+s)}}
\\[0.5em]
&\times
Y_*^{\,2-\frac{2}{c+d(-1+s)}-\frac{2}{-1+s}}\,
(1+Y_*)^{\frac{2-2s}{c+d(-1+s)}}\,
(1-2s-sY_*)
\\[0.5em]
&\times
\Bigl(Y_*^{\frac{1}{-1+s}}+Y_*^{\frac{s}{-1+s}}\Bigr)\,
\mathcal{K}_1(Y_*)
\\[0.5em]
&\times
\Bigl(
2(1-2s)s
+Y_*\bigl(3+s(-5+2s-4s^{2})-s(-1+s+2s^{2})Y_*\bigr)
\Bigr)\,\phi_*,
\end{aligned}
\ee
\be
\begin{aligned}
\mathcal{F}_6(\phi_*,Y_*) \equiv\;&
2^{-\frac{s\bigl(c+(-2+d)(-1+s)\bigr)}{\bigl(c+d(-1+s)\bigr)(-1+s)}}\,
c^{2}\,
\gamma^{\frac{2(-1+s)}{c+d(-1+s)}}\,
\bigl(c+d(-1+s)\bigr)\,s\,
(-1+2s)^{-1+\frac{2}{c+d(-1+s)}}
\\[0.5em]
&\times
Y_*^{\,2-\frac{2}{c+d(-1+s)}-\frac{2}{-1+s}}\,
(1+Y_*)^{\frac{2-2s}{c+d(-1+s)}}\,
(1+sY_*)
\\[0.5em]
&\times
\Bigl(Y_*^{\frac{1}{-1+s}}+Y_*^{\frac{s}{-1+s}}\Bigr)\,
\mathcal{K}_1(Y_*)
\\[0.5em]
&\times
\Bigl(
2(1-2s)s
+Y_*\bigl(3+s(-5+2s-4s^{2})-s(-1+s+2s^{2})Y_*\bigr)
\Bigr)\,\phi_*,
\end{aligned}
\ee
\be
\begin{aligned}
\mathcal{F}_7(\phi_*,Y_*) \equiv\;&
2^{-\frac{s\bigl(c+(-2+d)(-1+s)\bigr)}{\bigl(c+d(-1+s)\bigr)(-1+s)}}\,
c^{2}\,
\gamma^{\frac{2(-1+s)}{c+d(-1+s)}}\,
\bigl(c+d(-1+s)\bigr)\,
(-1+2s)^{-1+\frac{2}{c+d(-1+s)}}
\\[0.5em]
&\times
Y_*^{\,1-\frac{2}{c+d(-1+s)}-\frac{2}{-1+s}}\,
(1+Y_*)^{\frac{2-2s}{c+d(-1+s)}}\,
(1-2s-sY_*)\,(1+sY_*)
\\[0.5em]
&\times
\Bigl(Y_*^{\frac{1}{-1+s}}+Y_*^{\frac{s}{-1+s}}\Bigr)\,
\mathcal{K}_1(Y_*)
\\[0.5em]
&\times
\Bigl(
2(1-2s)s
+Y_*\bigl(3+s(-5+2s-4s^{2})-s(-1+s+2s^{2})Y_*\bigr)
\Bigr)\,\phi_*,
\end{aligned}
\ee
\be
\begin{aligned}
\mathcal{F}_8(\phi_*,Y_*) \equiv\;&
-\,2^{\frac{c+d(-1+s)+s}{c+d(-1+s)}}\,
(d-1)d\,
\gamma^{\frac{c+(1+d)(-1+s)}{c+d(-1+s)}}\,
\bigl(c+(-2+d)(-1+s)\bigr)\,
(-1+2s)^{\frac{1}{c+d(-1+s)}}
\\[0.5em]
&\times
Y_*^{\frac{1}{1-s}+\frac{1}{-c+d-ds}}\,
(1+Y_*)^{\frac{1-s}{c+d(-1+s)}}\,
(1+sY_*)^{3}\,
(-1+2s+sY_*)^{2}\,
\mathcal{K}_0(Y_*)\;
\phi_*^{\,d+\frac{c}{-1+s}},
\end{aligned}
\ee
\be
\begin{aligned}
\mathcal{F}_9(\phi_*,Y_*) \equiv\;&
2^{\frac{c(-3+s)+(-1+s)\bigl(d(-3+s)+s\bigr)}
{\bigl(c+d(-1+s)\bigr)(-1+s)}}\,
c d\,
\gamma^{\frac{c+(1+d)(-1+s)}{c+d(-1+s)}}\,
\bigl(c+(-2+d)(-1+s)\bigr)
\\[0.5em]
&\times
(-1+2s)^{\frac{1}{c+d(-1+s)}-\frac{s}{-1+s}}\,
(-2+4s)^{\frac{2}{-1+s}}\,
Y_*^{-\frac{4}{-1+s}+\frac{s}{-1+s}+\frac{1}{-c+d-ds}}
\\[0.5em]
&\times
(1+Y_*)^{-\frac{c+(1+d)(-1+s)}{c+d(-1+s)}}\,
(1+sY_*)^{2}\,
(-1+2s+sY_*)\,
\bigl(-3+7s-2s^{2}+s(-1+3s)Y_*\bigr)
\\[0.5em]
&\times
\Bigl(Y_*^{\frac{1}{-1+s}}+Y_*^{\frac{s}{-1+s}}\Bigr)\,
\mathcal{K}_1(Y_*)\;
\phi_*^{\,d+\frac{c}{-1+s}},
\end{aligned}
\ee
\be
\begin{aligned}
\mathcal{F}_{10}(\phi_*,Y_*) \equiv\;&
2^{\frac{c(-2+s)+(-1+s)\bigl(d(-2+s)+s\bigr)}
{\bigl(c+d(-1+s)\bigr)(-1+s)}}\,
\gamma^{\frac{2c+(1+2d)(-1+s)}{c+d(-1+s)}}\,
\bigl(c+(-1+d)(-1+s)\bigr)\,
(-1+2s)^{\frac{1}{c+d(-1+s)}}
\\[0.5em]
&\times
Y_*^{-\frac{4}{-1+s}+\frac{1}{-c+d-ds}}\,
(1+Y_*)^{-\frac{c+(1+d)(-1+s)}{c+d(-1+s)}}\,
(-1+2s+sY_*)\,(d+d s Y_*)^{2}
\\[0.5em]
&\times
\Bigl(
3\,2^{\frac{1+s}{-1+s}}(-1+2s)^{\frac{s}{-1+s}}
+s(-2+4s)^{\frac{1}{-1+s}}
\bigl(-2^{\frac{s}{-1+s}}+2^{3+\frac{1}{-1+s}}s\bigr)Y_*
\Bigr)
\\[0.5em]
&\times
\Bigl(Y_*^{\frac{1}{-1+s}}+Y_*^{\frac{s}{-1+s}}\Bigr)\,
\mathcal{K}_2(Y_*)\;
\phi_*^{\,2\left(d+\frac{c}{-1+s}\right)},
\end{aligned}
\ee
\be
\begin{aligned}
\mathcal{F}_{11}(\phi_*,Y_*) \equiv\;&
2^{\frac{c+d(-1+s)+2s}{c+d(-1+s)}}\,
(d-1)d\,
\gamma^{\frac{c+(2+d)(-1+s)}{c+d(-1+s)}}\,
\bigl(c+d(-1+s)\bigr)\,
(-1+2s)^{\frac{2}{c+d(-1+s)}}
\\[0.5em]
&\times
Y_*^{-\frac{2}{c+d(-1+s)}-\frac{3}{-1+s}}\,
(1+Y_*)^{\frac{2-2s}{c+d(-1+s)}}\,
(1+sY_*)^{2}\,
(-1+2s+sY_*)^{2}
\\[0.5em]
&\times
\Bigl(Y_*^{\frac{1}{-1+s}}+Y_*^{\frac{s}{-1+s}}\Bigr)\,
\mathcal{K}_2(Y_*)\;
\phi_*^{\,1+d+\frac{c}{-1+s}},
\end{aligned}
\ee
with
\bea
&& \mathcal{K}_0(Y_*) \equiv
d(-2+4s)^{\frac{s}{-1+s}}
+\Bigl(2^{\frac{s}{-1+s}}c\,(-1+2s)^{\frac{1}{-1+s}}
+d(-2+4s)^{\frac{s}{-1+s}}\Bigr)Y_*,
\\
&&\mathcal{K}_1(Y_*) \equiv
d(-1+2s)^{\frac{1}{1-s}}(-2+4s)^{\frac{s}{-1+s}}
\Bigl(Y_*^{\frac{1}{-1+s}}+Y_*^{\frac{s}{-1+s}}\Bigr)
+2^{\frac{s}{-1+s}}c\,Y_*^{\frac{s}{-1+s}},
\\
&& \mathcal{K}_2(Y_*) \equiv
d(-2+4s)^{\frac{s}{-1+s}}\,Y_*^{\frac{1}{-1+s}}
+\Bigl(2^{\frac{s}{-1+s}}c\,(-1+2s)^{\frac{1}{-1+s}}
+d(-2+4s)^{\frac{s}{-1+s}}\Bigr)\,Y_*^{\frac{s}{-1+s}}.
\eea

\end{appendix}

\bibliography{bio}

\end{document}